\documentclass[11pt,a4paper]{elsarticle}

\usepackage{amsmath}
\usepackage{graphicx}
\usepackage{bm}
\usepackage{xcolor}
\usepackage{algorithmic}
\usepackage{subfig}
\usepackage{verbatim}
\usepackage{MnSymbol}%
\usepackage{wasysym}%
\usepackage[colorinlistoftodos]{todonotes}
\usepackage[colorlinks=true, allcolors=blue]{hyperref}
\usepackage[section]{placeins}
\usepackage[margin=2.5cm]{geometry}
\usepackage{tikz}
\usepackage{xcolor}

\makeatletter
\newcommand{\btriangle}{\mathpalette\btriangle@\relax}
\newcommand{\btriangle@}[2]{%
  \begingroup
  \sbox\z@{$\m@th#1\triangle$}%
  \makebox[\wd\z@]{%
    \raisebox{0.04\height}{%
      \resizebox{1.1\wd\z@}{0.96\ht\z@}{%
        $\m@th#1\blacktriangle$%
      }%
    }%
  }%
  \endgroup
}
\makeatother

\makeatletter
\newcommand\bigDiamond{\mathop{\mathpalette\bigDi@mond\relax}}
\newcommand\bigDi@mond[2]{%
  \vcenter{\hbox{\m@th
    \scalebox{\ifx#1\displaystyle 2\else1.2\fi}{$#1\Diamond$}%
  }}%
}
\newcommand\bigLozenge{\mathop{\mathpalette\bigL@zenge\relax}}
\newcommand\bigL@zenge[2]{%
  \vcenter{\hbox{\m@th
    \scalebox{\ifx#1\displaystyle 2\else1.2\fi}{$#1\blacklozenge$}%
  }}%
}
\newcommand{\tikzcircle}[2][black,fill=black]{\tikz[baseline=-0.5ex]\draw[#1,radius=#2] (0,0) circle ;}%

\makeatother

\journal{Journal of Computational Physics}
\begin{document}
\begin{frontmatter}
\title{Accelerating the force-coupling method for hydrodynamic interactions in periodic domains}

\author[mymainaddress]{Hang Su}

\author[mymainaddress]{Eric E Keaveny}

\address[mymainaddress]{Department of Mathematics, Imperial College London, South Kensington Campus, London, SW7 2AZ, UK}

\begin{abstract}
The efficient simulation of fluid-structure interactions at zero Reynolds number requires the use of fast summation techniques in order to rapidly compute the long-ranged hydrodynamic interactions between the structures.  One approach for periodic domains involves utilising a compact or exponentially decaying kernel function to spread the force on the structure to a regular grid where the resulting flow and interactions can be computed efficiently using an FFT-based solver.  A limitation to this approach is that the grid spacing must be chosen to resolve the kernel and thus, these methods can become inefficient when the separation between the structures is large compared to the kernel width.  In this paper, we address this issue for the force-coupling method (FCM) by introducing a modified kernel that can be resolved on a much coarser grid, and subsequently correcting the resulting interactions in a pairwise fashion.  The modified kernel is constructed to ensure rapid convergence to the exact hydrodynamic interactions and a positive-splitting of the associated mobility matrix.  We provide a detailed computational study of the methodology and establish the optimal choice of the modified kernel width, which we show plays a similar role to the splitting parameter in Ewald summation.  Finally, we perform example simulations of rod sedimentation and active filament coordination to demonstrate the performance of fast FCM in application.

\end{abstract}
\end{frontmatter}


\section{Introduction}
Micron-scale fluid-structure interactions are present throughout a range of industrial processes involving colloidal particles and suspensions, as well as natural processes including those in cell-level biology.  Notable examples from biology include the movement of fluid driven by flagella and cilia \cite{lauga_hydrodynamics_2009,elgeti_physics_2015,brennen_fluid_1977}, the flexible and motile hair-like organelles protruding from cell surfaces, and the dynamical rearrangement of filament-motor protein complexes in the cell interior \cite{shelley2016dynamics}.  Many interesting examples arise also in engineering applications, including the rheological changes exhibited by flowing suspensions of particles and fibres \cite{mueller2010rheology,stickel2005fluid,derakhshandeh_rheology_2011,du_roure_dynamics_2017}, the self-assembly of structures in more advanced materials such as magnetorheological fluids \cite{de2011magnetorheological}, and the motion and interactions of exotic colloidal particles such as nanomotors and other types of phoretic particles \cite{moran2017phoretic}.  A unifying feature of these systems is the presence of long-ranged hydrodynamic interactions that couple the motion of structures that are present.  

For microscopic objects moving through viscous fluids, viscous forces typically dominate over inertia resulting in the fluid motion being governed by the linear and steady Stokes equations.  This, along with the negligible effect of structure inertia, confers a linear relationship between the forces and velocities governing the motion of the structures.  The matrix relating the velocities and forces is called the mobility matrix.  While the relationship is linear, the mobility matrix is itself configuration dependent and dense due to the slow decay of the fluid velocity fields generated by the forced structures.  In direct implementations of methods for particulate suspensions such as Brownian \cite{ermak1978brownian,graham2018microhydrodynamics} and Stokesian dynamics \cite{brady1988stokesian}, the application of the mobility matrix is performed by pairwise summation -- an $O(N^2)$ calculation where $N$ is the number of degrees of freedom.  This scaling for the computational cost is also present in direct implementations of other approaches such as the boundary element method \cite{pozrikidis1992boundary}, the rigid multiblob method \cite{balboa2017hydrodynamics}, or the method of regularised Stokeslets \cite{cortez2001method,cortez2005method}, where multiple degrees of freedom are associated with each structure.  Thus, for simulations involving many degrees of freedom, computational costs quickly grow prohibitive and approaches that circumvent pairwise summation need to be considered.

Reducing the computational cost often requires taking advantage of fast summation techniques, such as the fast multipole method \cite{greengard1987fast,tornberg2008fast}, that provide the action of the mobility matrix without ever computing the mobility matrix directly.  For periodic domains, such methods can be constructed around the fast Fourier transform (FFT).  In this context, there are two related approaches.  The first stems directly from classical Ewald summation \cite{hasimoto1959periodic} where the application of the inverse Stokes operator is split between sums in real and Fourier space.  By introducing an appropriately chosen splitting function, rapid convergence of both the sums can be ensured.  The sum in real space can be interpreted as a local, pairwise correction to the changes to the mobility matrix introduced by the splitting function.  In terms of computation, the sum in Fourier space can be performed for all degrees of freedom simultaneously, while the pairwise correction only needs to be performed for degrees of freedom in close proximity.  This approach has been applied successfully for the evaluation of the interactions based on point-forces (Stokeslets) \cite{lindbo2010spectrally}, as well as the Rotne Prager Yamakawa (RPY) tensor in the positively split Ewald (PSE) method \cite{fiore2017rapid} and incorporated into accelerated \cite{sierou2001accelerated} and fast \cite{fiore2019fast} Stokesian Dynamics.

The second approach is to instead consider regular, localised force distributions that can be evaluated on a grid where the Stokes equations can be solved using an FFT-based method.  This is done, for example with the immersed boundary method (IBM) \cite{peskin_immersed_2002,bringley2008validation} and the force-coupling method (FCM) \cite{maxey_localized_2001,lomholt_force-coupling_2003,yeo_simulation_2010} with a main difference between the two being the particular choice of function, or kernel, used to transfer the structure force to the grid.  Additionally, with these methods structure velocities are obtained by using the same kernels to interpolate, or volume average, the resulting fluid flow.  The result is a positive definite mobility matrix and proper energy balance with the viscous dissipation in the surrounding fluid.  As discussed in \cite{fiore2017rapid}, a limitation of this approach is that the grid must be chosen to provide sufficient resolution of the kernel.  As a result, computations can become expensive in cases where the separation of the structures is large compared to the kernel width.  To emphasise this point, \cite{fiore2017rapid} showed that PSE utilises a maximum wavenumber approximately 3 times smaller than that needed for FCM.  

In this paper, we develop a fast implementation of FCM which alleviates this limitation by eliminating the need for the grid to resolve the FCM kernel.  We accomplish this by substituting the FCM kernel by a modified kernel with a larger width, and correcting the resulting particle velocities to ensure errors are below a user specified tolerance.  This extends similar ideas developed in \cite{graham2018microhydrodynamics,hernandez2007fast} for Stokeslet interactions, where regularised forces in real space, rather than an Ewald splitting function in Fourier space, is used to formulate the Ewald summation.  Our modified kernel is carefully chosen to ensure that the resulting pairwise mobility converges exponentially to the standard FCM mobility as particle separation distance increases.  Additionally, the specific choice of modified kernel ensures positive splitting of the force-velocity mobility matrix, resulting in SPD matrices for both the modified mobility matrix as well as the local pairwise correction matrix.  We also extend fast FCM to allow for torques and rotations.  We present a GPU-based implementation of fast FCM and perform a number of tests to demonstrate how parameters can be tuned to optimise computational performance.  In doing so, we find that in many cases fast FCM can be an order of magnitude faster than the standard FCM computation.  We provide example simulations of rod sedimentation and active filament dynamics that demonstrate the effectiveness of fast FCM in application and as part of larger computations.

\section{Force-coupling method: the force-velocity mobility matrix} \label{sec:FCM}
We begin by reviewing FCM \cite{maxey_localized_2001} for the motion of $N$ particles of hydrodynamic radius $a$ through a fluid with viscosity $\eta$ whose flow is governed by the Stokes equations.  Each particle $n = 1,\dots, N$ is centred at $\bm{Y}_n$ and exerts the force $\bm{F}_n$ on the fluid.  The fluid and particles occupy domain $\Omega$.  The force on each particle $n$ is transferred to the fluid using a Gaussian distribution, or kernel, 
\begin{align}
    \Delta_{n}(\bm{x};\sigma) = (2\pi\sigma^2)^{-3/2}\exp (-|\bm{x}-\bm{Y}_n|^2/2\sigma^2).
\end{align}
Accordingly, the Stokes system for the resulting fluid flow $\bm{u}(\bm{x})$ and pressure $p(\bm{x})$ is,
\begin{align}
    -\eta \nabla^2 \bm{u} + \bm{\nabla} p &= \mathcal{J}^{\dagger}\left[ \mathcal{F} \right] \nonumber \\
    \bm{\nabla} \cdot \bm{u} &=0, \label{eq:FCMstokes_system}
\end{align}
for $\bm{x}\in \Omega$, where $\mathcal{F} = \left[\bm{F}_1^\dagger, \bm{F}_2^\dagger, \dots, \bm{F}_N^\dagger \right]^\dagger$ is the $3N \times 1$ vector containing the components of the forces on all particles and $\mathcal{J}^{\dagger}\left[ \cdot \right]$ is the spreading operator that transfers the forces on all particles to the fluid such that,
\begin{align}
    \mathcal{J}^{\dagger}\left[ \mathcal{F} \right] = \sum^{N}_{n=1} \bm{F}_{n} \Delta_{n}(\bm{x};\sigma).
\end{align}
The motion of the particles is determined using the kernel to locally average $\bm{u}(\bm{x})$.  The velocity of particle $m$ is then given by,
\begin{align}
    \bm{V}_m &= (\mathcal{J}\left[ \bm{u} \right])_{m} = \int_{\Omega} \bm{u}(\bm{x})\Delta_{m}(\bm{x}; \sigma) d^3\bm{x} \label{eq:avg_particle_vel},
\end{align}
recognising also that the interpolation operator $\mathcal{J}$ and the spreading operator are adjoints.  The kernel width, $\sigma$, sets the hydrodynamic radius of the particles via $a = \sigma \sqrt{\pi}$.  This particular choice of $a$ recovers the Stokes drag law for a single particle in an unbounded domain \cite{maxey_localized_2001}.  

The successive actions of spreading the particle force to the fluid, solving for the resulting fluid velocity, and interpolating the fluid velocity to obtain the particle velocities provides the single linear relationship between the particle forces, $\mathcal{F}$, and velocities, and $\mathcal{V} = \left[\bm{V}_1^\dagger, \bm{V}_2^\dagger, \dots, \bm{V}_N^\dagger \right]^\dagger$,
\begin{align}
    \mathcal{V} = \mathcal{M}^{\mathcal{V}\mathcal{F}} \mathcal{F}
\end{align}
where $\mathcal{M}^{\mathcal{V}\mathcal{F}}$ is the $3N \times 3N$ force-velocity mobility matrix.  This can be written in terms of the operator $\mathcal{J}$ and $\mathcal{L}^{-1}$, the inverse of Stokes operator, as 
\begin{align}
    \mathcal{M}^{\mathcal{V}\mathcal{F}}[\cdot] &= \mathcal{J}[\mathcal{L}^{-1}[\mathcal{J}^{\dagger}[\cdot]]]. \label{eq:MVF_operators}
\end{align}
\subsection{Grid-based algorithm for FCM in periodic domains}\label{sec:gridFCM}

For simulations of particles in periodic domains where $\Omega = [0, L_x)\times [0, L_y) \times [0, L_z)$, the FCM mobility matrix can be applied using an FFT-based algorithm described previously in \cite{yeo_simulation_2010,keaveny2014fluctuating}.  Here, $\Omega$ is discretised using a uniform grid with spacing $\Delta x$ and size $M_x\times M_y \times M_z$, where $M_x$, $M_y$ and $M_z$ are the number of gridpoints in the $x-$, $y-$, and $z-$directions, respectively.  The total number of grid points is then $M = M_xM_yM_z$.  The algorithm consists of three main steps, each corresponding to the action of one of the continuous operators described above.

\begin{enumerate}
\item \label{spread} \textbf{Applying $\mathcal{J}^\dagger$}: The particle force is communicated to the fluid by evaluating the Gaussian force distributions on the grid.   Although the Gaussian is not compactly supported, it decays rapidly and the tails can be safely truncated in an error-controlled fashion.  Thus, each Gaussian is supported locally on the grid by $M_G \times M_G \times M_G$ gridpoints. 
\item \label{solve} \textbf{Applying the inverse Stokes operation, $\mathcal{L}^{-1}$}: 
\begin{enumerate}
\item The Fourier transform of the total force distribution on the grid is computed using the FFT.
\item The fluid velocity in Fourier space is obtained by applying the inverse Stokes operator in Fourier space.
\item The inverse Fourier transform of the fluid velocity is computed using the FFT.
\end{enumerate}
\item \label{interpolate} \textbf{Applying $\mathcal{J}$}: The particle velocities are obtained by evaluating numerically the integral appearing in \eqref{eq:avg_particle_vel} using the trapezoidal rule.  Again, the sums are performed using the truncated Gaussians on grid of size $M_G \times M_G \times M_G$.
\end{enumerate}
The operation count associated with evaluating the force on the grid and interpolating the resulting flow is $\mathcal{O}(N M_{G}^3)$, while that associated with the application of the inverse Stokes operator is $\mathcal{O}(M \log M)$. While we see that this algorithm avoids the $\mathcal{O}(N^2)$ scaling associated with pairwise computation, when the grid is large compared to the number of particles, the cost of the FCM algorithm will exceed that of pairwise evaluation.  

The accuracy of this algorithm hinges on the grid being sufficiently small as to resolve the Gaussian kernel used in $\mathcal{J}$ and $\mathcal{J}^{\dagger}$, i.e. $\Delta x < \sigma$.  Thus, the grid spacing, and hence number of grid points, is determined by $\sigma$ and consequently the hydrodynamic radius, $a$.  As a result, for simulations at low volume fraction where the separation between particles is large compared to the particle size, the FFT and applying the inverse Stokes operator will require excessive computational time.  It is also worth noting that the grid-based approach can incur a high memory cost. Flow data on a $1024\times1024\times1024$ grid using double precision requires 25.8GB RAM.  

The purpose of this paper is to formulate fast FCM by decoupling the grid spacing from the Gaussian width to facilitate more efficient computation across all particle volume fractions.  We accomplish this by replacing the Gaussian kernel in the FCM algorithm by a modified kernel of larger width, and then correcting the resulting particle velocities by a pairwise computation.  An essential aspect of this procedure is constructing a modified kernel that ensures the number of particle pairs requiring correction remains small while simultaneously having the kernel width as large as possible.

\subsection{FCM pairwise mobility matrix}\label{section:flow_response}
A key piece of information needed to formulate fast FCM is the FCM pairwise mobility matrix, $\bm{M}^{\bm{VF}}_{nm}$, which provides the contribution to the velocity of particle $n$ due to the force on particle $m$.  We can obtain an analytical expression for $\bm{M}^{\bm{VF}}_{nm}$ using the expressions found in \cite{maxey_localized_2001} for the fluid velocity generated by a Gaussian force distribution.  

\textcolor{blue}{
The flow generated by $\bm{f}(\bm{x}) = \bm{F}\Delta(\bm{x};\sigma)$ with $\Delta(\bm{x};\sigma) = (2\pi\sigma^2)^{-3/2}\exp (-r^2/2\sigma^2)$ and $r=\lVert \bm{x}\rVert$
}can be expressed as 
\begin{align} \label{eq:FCMflow}
\bm{u}(\bm{x}) = \bm{S}(\bm{x};\sigma)\bm{F}.
\end{align}
where we can write $\bm{S}(\bm{x};\sigma) = \bm{S}^{(1)}(\bm{x};\sigma) + \bm{S}^{(2)}(\bm{x};\sigma) + \bm{S}^{(3)}(\bm{x};\sigma)$ with
\begin{align}
\bm{S}^{(1)}(\bm{x};\sigma) &= \frac{1}{8\pi \eta r}\left(\bm{I} + \frac{\bm{x}\bm{x}^T}{r^2}\right)\textrm{erf}\left(\frac{r}{\sigma\sqrt{2}}\right), \label{eq:S1} \\
\bm{S}^{(2)}(\bm{x};\sigma) &= \frac{1}{8\pi \eta r^3}\left(\bm{I} - 3\frac{\bm{x}\bm{x}^T}{r^2}\right)\sigma^2\textrm{erf}\left(\frac{r}{\sigma\sqrt{2}}\right), \label{eq:S2} \\
\bm{S}^{(3)}(\bm{x};\sigma) &= -\frac{\sigma^2}{2 \eta}\left(\bm{I} - 3\frac{\bm{x}\bm{x}^T}{r^2}\right)\left(\frac{\sigma^2}{r^2}\right)\Delta\left(\bm{x};\sigma \right). \label{eq:S3}
\end{align}
In terms of the Oseen tensor,
\begin{align}
\bm{G}(\bm{x}) = \frac{1}{8\pi \eta r}\left(\bm{I} + \frac{\bm{x}\bm{x}^T}{r^2}\right),
\end{align}
we have 
\begin{align}
\bm{S}^{(1)}(\bm{x};\sigma) &= \textrm{erf}\left(\frac{r}{\sigma\sqrt{2}}\right)\bm{G}(\bm{x}),  \\
\bm{S}^{(2)}(\bm{x};\sigma) &= \frac{\sigma^2}{2} \textrm{erf}\left(\frac{r}{\sigma\sqrt{2}}\right)\nabla^2 \bm{G}(\bm{x})
\end{align}

The expression for the pairwise mobility matrix for FCM follows directly from that for the flow.  Namely, the entries of the mobility matrix that relate the velocity of particle $n$ to the force on particle $m$ are
\begin{align}
    \bm{M}^{\bm{VF}}_{nm} & = \bm{S}(\bm{Y}_n - \bm{Y}_m;\sigma\sqrt{2}).\label{eq:M_VF}
\end{align}

We see that this is identical to the expression for the flow field, but with the envelope size replaced by \textcolor{blue}{$\sigma \sqrt{2}$}.  This slight modification is a result of the volume averaging.  The validity of this expression is most easily established using Fourier integrals, as shown in \ref{appendix:fcm}.

Along with these expressions, it will also be useful to have at hand the flow generated by $\bm{f}(\bm{x}) = \bm{H}\nabla^2\Delta(\bm{x};\sigma)$, which is 
\begin{align}
    \bm{u}(\bm{x}) = \bm{Q}(\bm{x};\sigma)\bm{H},
\end{align}
where $\bm{Q}(\bm{x};\sigma)$ can be decomposed into two terms $\bm{Q}(\bm{x};\sigma) = \bm{Q}^{(1)}(\bm{x};\sigma) + \bm{Q}^{(2)}(\bm{x};\sigma)$ where
\textcolor{blue}{
\begin{align}
    \bm{Q}^{(1)}(\bm{x};\sigma)&= \frac{1}{4\pi\eta r^3}\left(\bm{I} - 3\frac{\bm{x}\bm{x}^T}{r^2} \right) \textrm{erf}\left(\frac{r}{\sigma\sqrt{2}}\right), \label{eq:Q1}\\
    \bm{Q}^{(2)}(\bm{x};\sigma)&=-\frac{1}{\eta}\left( (1+\frac{\sigma^2}{r^2})\bm{I} - (1+\frac{3\sigma^2}{r^2})\frac{\bm{x}\bm{x}^T}{r^2} \right) \Delta\left(\bm{x};\sigma \right).
    \label{eq:Q2}
\end{align}
}
The expression \eqref{eq:Q1} can also be written as
\begin{align}
    \bm{Q}^{(1)}(\bm{x};\sigma) &= \textrm{erf}\left(\frac{r}{\sigma\sqrt{2}}\right) \nabla^2 \bm{G}(\bm{x}).
\end{align}

\section{Fast FCM}\label{sec:fastfcm}
With the results from the previous sections established, we are in the position to formulate and justify the fast FCM framework.  As is typically done in Ewald splitting, we decompose the mobility matrix into two parts
\begin{align}
    \mathcal{M^{VF}} = \widetilde{\mathcal{M^{VF}}} + (\mathcal{M^{VF}} - \widetilde{\mathcal{M^{VF}}}), \label{eq:ewald}
\end{align}
and aim for the action of $\widetilde{\mathcal{M^{VF}}}$ to be evaluated efficiently using a grid-based computation and the correction $(\widetilde{\mathcal{M^{VF}}} - \mathcal{M^{VF}})$ to be applied pairwise but only for a limited number of pairs whose separations are within a cut-off radius, i.e. $(\widetilde{\mathcal{M^{VF}}} - \mathcal{M^{VF}})$ is sparse.

In standard Ewald splitting, the decomposition and aims are achieved by introducing the splitting function $H(k;\xi)$, where $k = |\bm{k}|$ is the magnitude of the wavenumber $\bm{k}$ and $\xi$ is the splitting parameter, in the Fourier transform of the inverse Stokes operator such that $\hat{\mathcal{L}}^{-1} = \hat{\mathcal{L}}^{-1}H(k;\xi) + \hat{\mathcal{L}}^{-1}(1 - H(k;\xi))$, which are then associated with $\widetilde{\mathcal{M^{VF}}}$ and $(\mathcal{M^{VF}} - \widetilde{\mathcal{M^{VF}}})$, respectively.  The splitting function is selected to decay exponentially with increasing $k$, with the decay rate controlled by the splitting parameter, $\xi$.   Common choices for $H(k;\xi)$ include the Hasimoto \cite{hasimoto1959periodic} 
\begin{align}
H_{H}(k; \xi) = \left(1 + \frac{k^2}{4\xi^2}\right)e^{-k^2/4\xi^2}
\end{align}
and Beenakker \cite{beenakker1986ewald}
\begin{align}
H_{B}(k; \xi) = \left(1 + \frac{k^2}{4\xi^2} + \frac{k^4}{8\xi^4}\right)e^{-k^2/4\xi^2}
\end{align}
splitting functions.  

With fast FCM, we achieve a similar splitting by replacing the kernel $\Delta(\bm{x}; \sigma)$ in FCM with a modified kernel $\widetilde{\Delta}(\bm{x}; \Sigma)$, which we use to compute the action of the approximate mobility $\widetilde{\mathcal{M}^{VF}}$ using the standard, FFT-based FCM algorithm.  Thus, to ensure that the cost of applying $\widetilde{\mathcal{M}^{VF}}$ will be less then that of $\mathcal{M}^{VF}$, we must have $\Sigma > \sigma$ to be able to use a smaller grid.  At the same time, however, the choice of $\widetilde{\Delta}(\bm{x}; \Sigma)$ should yield an approximate pairwise mobility matrix that satisfies $\lVert \widetilde{\bm{M}^{\bm{VF}}_{nm}} - \bm{M}^{\bm{VF}}_{nm} \rVert \rightarrow 0$ exponentially as $\lVert \bm{Y}_n - \bm{Y}_m \rVert \rightarrow \infty$.  This requirement ensures that the correction matrix $\widetilde{\mathcal{M^{VF}}} - \mathcal{M^{VF}}$ is sparse and the number of pairwise corrections needed to achieve a given tolerance is minimised.

With these requirements in mind, we utilise the modified kernel, 
\begin{align}
    \widetilde{\Delta}_{n}(\bm{x}; \Sigma) = \left(1 + \frac{\sigma^2 - \Sigma^2}{2}\nabla^2 \right) \Delta_{n}(\bm{x};\Sigma).
\end{align}
with $\Sigma > \sigma$.  The inclusion of the operator $\left(1 + (\sigma^2 - \Sigma^2)/2\right)\nabla^2$, as we will see, enables the exponential convergence of the pairwise mobility.

\subsection{The action of $\widetilde{\mathcal{M^{VF}}}$}
Computing the action $\widetilde{\mathcal{M^{VF}}}$ follows the same steps as the standard FCM computation described in Section \ref{sec:FCM} with $\Delta_{n}(\bm{x}; \sigma)$ replaced by $\widetilde{\Delta}_{n}(\bm{x}; \Sigma)$.  Accordingly, we consider the Stokes system
\begin{align}
    -\eta \nabla^2 \widetilde{\bm{u}} + \nabla \widetilde{p} &= \widetilde{\mathcal{J}}^{\dagger}[\mathcal{F}], \\
    \bm{\nabla} \cdot \widetilde{\bm{u}} &=0,
\end{align}
where
\begin{align}
    \widetilde{\mathcal{J}}^{\dagger}[\mathcal{F}] = \sum^{N}_{n=1} \bm{F}_n \widetilde{\Delta}_{n}(\bm{x};\Sigma) \label{eq:force_kernel_mod}, 
\end{align}
and compute the flow field.  Then, from the fluid velocity, we determine the approximate particle velocities by evaluating 
\begin{align}
    \widetilde{\bm{V}}_m = (\widetilde{\mathcal{J}}\left[ \bm{u} \right])_{m} = \int_{-\infty}^{\infty} \bm{u}(\bm{x}) \widetilde{\Delta}_m(\bm{x}; \Sigma) d^3\bm{x} \label{eq:avg_particle_vel_mod},
\end{align}

\subsection{The action of $\mathcal{M^{VF}} - \widetilde{\mathcal{M^{VF}}}$}
To correct these velocities, we utilise an analytical expression for the pairwise mobility.  We derive this expression from the integral representation \cite{keaveny2014fluctuating} of the pairwise mobility,
\begin{align}
    \widetilde{\bm{M}^{\bm{VF}}_{nm}} &= \int \int \widetilde{\Delta}(\bm{y}-\bm{Y}_{n}; \Sigma) \widetilde{\Delta}(\bm{x} - \bm{Y}_{m}; \Sigma) \bm{G}(\bm{x}-\bm{y}) d^3\bm{y}d^3\bm{x}, 
\end{align}
which we then expand as
\begin{align}
\begin{split}
    \widetilde{\bm{M}^{\bm{VF}}_{nm}} =& \int \int \Delta(\bm{y}-\bm{Y}_{n}; \Sigma) \Delta(\bm{x} - \bm{Y}_{m}; \Sigma)
    \times \bm{G}(\bm{x}-\bm{y}) d^3\bm{y}d^3\bm{x} \\
    &+ \int \int \Delta(\bm{y}-\bm{Y}_{n}; \Sigma) \Delta(\bm{x} - \bm{Y}_{m}; \Sigma)(\sigma^2-\Sigma^2)\nabla^2 \bm{G}(\bm{x}-\bm{y}) d^3\bm{y}d^3\bm{x} \\
    &+ \int \int \Delta(\bm{y}-\bm{Y}_{n}; \Sigma) \Delta(\bm{x} - \bm{Y}_{m}; \Sigma) \frac{(\sigma^2-\Sigma^2)^2}{4} \nabla^2 \nabla^2 \bm{G}(\bm{x}-\bm{y}) d^3\bm{y}d^3\bm{x}
\end{split}
\end{align}
Each integral can be related to the expressions for the FCM mobility matrices found in Section \ref{section:flow_response}.   Specifically, we have
\begin{align}
     \widetilde{\bm{M}^{\bm{VF}}_{nm}} &= \bm{S}(\bm{Y}_n - \bm{Y}_m; \sqrt{2}\Sigma) + (\sigma^2 - \Sigma^2)\bm{Q}(\bm{Y}_n - \bm{Y}_m; \sqrt{2}\Sigma) + \frac{(\sigma^2 - \Sigma^2)^2}{4} \bm{T}(\bm{Y}_n - \bm{Y}_m; \sqrt{2}\Sigma),
\end{align}
where $\bm{S}(\bm{x}; \sigma)$ and $\bm{Q}(\bm{x}; \sigma)$ are provided by \eqref{eq:S1}-\eqref{eq:S3} and \eqref{eq:Q1}-\eqref{eq:Q2}, respectively.  The last term, $\bm{T}(\bm{Y}_n - \bm{Y}_m; \sqrt{2}\Sigma)$, is new and given by  
\begin{align}
     \bm{T}(\bm{x}; \sigma) &= \frac{1}{\eta\sigma^2}\left(2\bm{I} + \frac{1}{\sigma^2}\left(\bm{x}\bm{x}^T - r^2\bm{I}\right)\right)\Delta(\bm{x};\sigma). 
\end{align}

Taking the difference with the FCM pairwise mobility \eqref{eq:M_VF}, we arrive at the expression for the correction matrix,
\begin{align}
\begin{split}
    \bm{M}^{\bm{VF}}_{nm} - \widetilde{\bm{M}^{\bm{VF}}_{nm}} =& \left(\bm{G}(\bm{Y}_n - \bm{Y}_m) + \sigma^2 \nabla^2 \bm{G}(\bm{Y}_n - \bm{Y}_m)\right) \left(\textrm{erf}\left(\frac{r}{\sigma\sqrt{2}}\right) - \textrm{erf}\left(\frac{r}{\Sigma\sqrt{2}}\right)\right) \\
&+ \bm{S}^{(3)}(\bm{Y}_n - \bm{Y}_m;\sigma\sqrt{2}) - \bm{S}^{(3)}(\bm{Y}_n - \bm{Y}_m;\Sigma\sqrt{2}) \\
& -\left( \sigma^2 - \Sigma^2 \right) \bm{Q}^{(2)}(\bm{Y}_n - \bm{Y}_m;\Sigma\sqrt{2}) - \frac{(\sigma^2 - \Sigma^2)^2}{4} \bm{T}(\bm{Y}_n - \bm{Y}_m; \Sigma \sqrt{2}). \label{eq:correction_VF}
\end{split}
\end{align}

Along with providing the expression needed to correct the particle velocities, \eqref{eq:correction_VF} reveals the important property that each term on the right-hand side decays exponentially as $\lVert \bm{Y}_n - \bm{Y}_m \rVert \rightarrow \infty$.  The expression for \textcolor{blue}{\eqref{eq:correction_VF} evaluated at $\lVert \bm{Y}_n - \bm{Y}_m \rVert = 0$, which is well-defined, is provided in \ref{appendix:selfcorrect} and is used to correct the self-mobility matrix for every particle.  Note also that we assume that the domain size $L$ is always sufficiently large so corrections are not needed between a particle and its own periodic images.}  As with standard Ewald splitting, rapid convergence is crucial to ensuring that the correction matrix is sparse.  We also observe that the parameter $\Sigma$ that controls the width of the modified kernel plays the same role as $\xi$ in Ewald splitting.  Accordingly, $\Sigma$ will need to be chosen to optimise the overall performance of fast FCM.  

\subsection{Positive splitting}
An additional desirable feature associated with this choice of $\widetilde{\Delta}$ is that it provides a positive splitting of the mobility matrix, namely that both $\widetilde{\mathcal{M^{VF}}}$ and $\mathcal{M^{VF}} - \widetilde{\mathcal{M^{VF}}}$ are symmetric positive definite matrices.  While positive splitting is not essential for the efficient evaluation of the hydrodynamic interactions, \textcolor{blue}{as discussed in} \cite{fiore2017rapid}, it is essential for the efficient computation of Brownian displacements when thermal fluctuations are to be included in the computation.  

First, we notice that $\widetilde{\mathcal{M^{VF}}}$ is positive definite by construction due to the positive definiteness of the inverse Stokes operator and the spreading operator being the adjoint of the interpolation operator.  It remains to show that $\mathcal{M^{VF}} - \widetilde{\mathcal{M^{VF}}}$ is positive definite. 

We begin with the expression for the pairwise FCM mobility matrix in terms of the Fourier transforms (see also \ref{appendix:fcm}) of $\Delta(\bm{x};\sigma)$ and the inverse Stokes operator, 
\begin{align}
    \bm{M^{VF}}_{nm} = (2\pi)^3\int e^{i\bm{k}\cdot(\bm{Y}_n - \bm{Y}_m )}\hat{\Delta}(\bm{k};\sigma) \hat{\mathcal{L}}^{-1}(\bm{k}) \hat{\Delta}(\bm{k};\sigma) d^3\bm{k},
\end{align}
where 
\begin{align}
   \hat{\Delta}(\bm{k};\sigma) &= (2\pi)^{-3}e^{-\sigma^2k^2/2}.
\end{align}
If we were to apply Ewald splitting directly, we would have 
\begin{align}
\begin{split}
    \bm{M^{VF}}_{nm} =& (2\pi)^3\int e^{i\bm{k}\cdot(\bm{Y}_n - \bm{Y}_m )}\hat{\Delta}(\bm{k};\sigma) \hat{\mathcal{L}}^{-1}(\bm{k})(1 - H(k;\xi)) \hat{\Delta}(\bm{k};\sigma) d^3\bm{k} \\
    &+ (2\pi)^3\int e^{i\bm{k}\cdot(\bm{Y}_n - \bm{Y}_m )}\hat{\Delta}(\bm{k};\sigma) \hat{\mathcal{L}}^{-1}(\bm{k})H(k;\xi) \hat{\Delta}(\bm{k};\sigma) d^3\bm{k}.
\end{split}
\end{align}
As described in \cite{fiore2017rapid}, the splitting is positive if $0 < H(k; \xi) < 1$ for $k > 0$.  To demonstrate that our modified kernel approach yields positive splitting, we find the corresponding splitting function and show that it satisfies this condition.  

Since,
\begin{align}
   \hat{\widetilde{\Delta}}(\bm{k};\Sigma) &= \left(1 - \frac{\sigma^2 - \Sigma^2}{2}k^2\right)(2\pi)^{-3}e^{-\Sigma^2k^2/2},
\end{align}
we can write
\begin{align}
    \hat{\widetilde{\Delta}}(\bm{k};\Sigma) &= \left(1 - \frac{\sigma^2 - \Sigma^2}{2}k^2\right)\left(e^{-\sigma^2k^2/2}\right)^{\Sigma^2/\sigma^2 - 1}\hat{\Delta}(\bm{k};\sigma).
\end{align}
This allows us to express $\widetilde{\bm{M}^{\bm{VF}}_{nm}}$ as,
\begin{align}
    \widetilde{\bm{M}^{\bm{VF}}_{nm}} = (2\pi)^3\int e^{i\bm{k}\cdot(\bm{Y}_n - \bm{Y}_m )}\hat{\Delta}(\bm{k};\sigma) \hat{\mathcal{L}}^{-1}(\bm{k})H_{FCM}(k;\Sigma,\sigma) \hat{\Delta}(\bm{k};\sigma) d^3\bm{k}.
\end{align}
where the fast FCM splitting function is
\textcolor{blue}{
\begin{align}
    H_{FCM}(k;\Sigma,\sigma) &= p(k)e^{-(\Sigma^2 - \sigma^2)k^2}, \\
    p(k;\Sigma,\sigma) &= 1 - (\sigma^2 - \Sigma^2)k^2 + \frac{(\sigma^2 - \Sigma^2)^2}{4}k^4.
\end{align}
}

We first notice that $H_{FCM}(k;\Sigma,\sigma) > 0$ which follows from $e^{-(\Sigma^2 - \sigma^2)k^2} > 0$ and as $\Sigma > \sigma$, $p(k;\Sigma,\sigma) = (1 - (\sigma^2 - \Sigma^2)k^2/2)^2 > 0$.  Differentiating $H_{FCM}(k;\Sigma,\sigma)$ and $p(k;\Sigma,\sigma)$, we see that 
\textcolor{blue}{
\begin{align}
    H'_{FCM}(k;\Sigma,\sigma) &= \left(p'(k;\Sigma,\sigma) - 2k p(k;\Sigma,\sigma) (\Sigma^2 - \sigma^2) \right)e^{-(\Sigma^2 - \sigma^2)k^2}, \\
    p'(k;\Sigma,\sigma) &= -2(\sigma^2 - \Sigma^2)k + (\sigma^2 - \Sigma^2)k^3.
\end{align}
}
and examining the expression for $H'(k;\Sigma,\sigma)$ more closely, we find
\begin{align}
    p'(k;\Sigma,\sigma) - 2k p(k;\Sigma,\sigma) (\Sigma^2 - \sigma^2) = k^3(\sigma^2 - \Sigma^2)^2\left[\frac{\sigma^2 - \Sigma^2}{2}k^2 - 1 \right].    
\end{align}
Thus, since $k\geq 0$ and $\Sigma > \sigma$, we have $p'(k;\Sigma,\sigma) - 2k p(k;\Sigma,\sigma) (\Sigma^2 - \sigma^2) \leq 0$, revealing that $H'_{FCM}(k;\Sigma,\sigma) \leq 0$ and $H_{FCM}(k;\Sigma,\sigma)$ is monotonically decreasing with the maximum $H_{FCM}(0;\Sigma,\sigma) = 1$.  Therefore, we have $0 < H_{FCM}(k;\Sigma,\sigma) < 1$ for $k > 0$, indicating that the splitting is positive.

From this analysis, we also see how the approach of using a modified kernel corresponds to the standard Ewald splitting framework, namely that $\hat{\widetilde{\Delta}}(\bm{k};\Sigma) = (H_{FCM}(k;\Sigma,\sigma))^{1/2} \hat{\Delta}(\bm{k};\sigma)$.  Thus, it would be possible to have used one of the standard splitting functions, such as that used by Hasimoto \cite{hasimoto1959periodic}, provided that $H_{H}(k;\xi)^{1/2}$ exists for all $k$ and its inverse Fourier transform can be found to evaluate its real-space representation on a grid.  It is important to note that while this direct correspondence is available for periodic domains, in other domains such as channels, using a modified kernel in real-space to perform the splitting, as done for Stokeslets \cite{graham2018microhydrodynamics,hernandez2007fast}, may be the only route to accelerating the computation.
\subsection{Extension of fast FCM to torques and angular velocities}
As introduced in \cite{lomholt_force-coupling_2003}, FCM can be extended to include torques and particle rotations.  Along with the force, $\bm{F}_n$, the torque, $\bm{T}_n$, on each particle $n$ is also spread to the fluid through
\begin{align}
    -\eta \nabla^2 \bm{u} + \bm{\nabla} p &= \mathcal{J}^{\dagger}\left[ \mathcal{F} \right] + 
    \mathcal{N}^{\dagger}\left[ \mathcal{T} \right], 
    \\
    \bm{\nabla}\cdot \bm{u} &=0, \label{eq:FCMstokes_system_full}
\end{align}
where $\mathcal{T} = \left[\bm{T}_1^\dagger, \bm{T}_2^\dagger, \dots, \bm{T}_N^\dagger \right]^\dagger$, $\mathcal{J}^{\dagger}$ is given in \eqref{eq:force_kernel_mod}, and 
\begin{align}
    \mathcal{N}^{\dagger}\left[ \mathcal{T} \right] = \frac{1}{2}\sum^{N}_{n=1} \bm{T}_n \times \nabla \Delta_{n}(\bm{x};\sigma_{D}).
\end{align}
After solving for the fluid flow, the angular velocity, $\bm{\Omega}_m$, on particle $m$ is found by applying $\mathcal{N}$ to the flow field, 
\begin{align}
    \bm{\Omega}_m &= (\mathcal{N}\left[ \mathbf{u} \right])_{m} = \frac{1}{2}\int_{-\infty}^{\infty} \nabla \times \mathbf{u}(\mathbf{x})\Delta_{m}(\mathbf{x}; \sigma_{D}) d^3\mathbf{x} \label{eq:avg_particle_rot}.
\end{align}
For particle rotations, the width of the Gaussian is $\sigma_{D} = a/(6\sqrt{\pi})^{1/3}$ to ensure FCM provides the correct value for the viscous torque on a single spherical particle.

With the addition of torques and angular velocities, $\mathcal{W} = \left[\bm{W}_1^\dagger, \bm{W}_2^\dagger, \dots, \bm{W}_N^\dagger \right]^\dagger$, the mobility relationship is now
\begin{align}
    \begin{pmatrix}
    \mathcal{V} \\
    \mathcal{W}
    \end{pmatrix}
    &= \mathcal{M}\begin{pmatrix}
    \mathcal{F} \\
    \mathcal{T}
    \end{pmatrix} =\begin{pmatrix}
    \mathcal{M}^\mathcal{VF} & \mathcal{M}^\mathcal{VT}  \\
    \mathcal{M}^\mathcal{WF} & \mathcal{M}^\mathcal{W T}
    \end{pmatrix}
    \begin{pmatrix}
    \mathcal{F} \\
    \mathcal{T}
    \end{pmatrix}, \label{eq:symmetric_FFT_mod}
\end{align}
where $\mathcal{M}^\mathcal{VF}$ is given by \eqref{eq:M_VF}, and the other mobility $3N\times 3N$ submatrices are
\begin{align}
    \mathcal{M}^{\mathcal{V}\mathcal{F}}[\cdot] &= \mathcal{J}[\mathcal{L}^{-1}[\mathcal{J}^{\dagger}[\cdot]]] \label{eq:MVF_operators_full} \\
    \mathcal{M}^{\mathcal{V}\mathcal{T}}[\cdot] &= \mathcal{J}[\mathcal{L}^{-1}[\mathcal{N}^{\dagger}[\cdot]]] \label{eq:MVT_operators} \\
    \mathcal{M}^{\mathcal{W}\mathcal{F}}[\cdot] &= \mathcal{N}[\mathcal{L}^{-1}[\mathcal{J}^{\dagger}[\cdot]]] \label{eq:MWF_operators} \\
    \mathcal{M}^{\mathcal{W}\mathcal{F}}[\cdot] &= \mathcal{N}[\mathcal{L}^{-1}[\mathcal{N}^{\dagger}[\cdot]]] \label{eq:MWT_operators}.
\end{align}

We may utilise the analytical expressions \cite{maxey_localized_2001,lomholt_force-coupling_2003} for the vorticity induced by single FCM force and torque distributions located at the origin to obtain expressions for the pairwise mobility matrices.  The vorticity resulting from a force distribution is,
\begin{align}
\bm{\omega}(\bm{x}) = \bm{R}(\bm{x};\sigma)\bm{F}.
\end{align}
where 
\begin{align}
\bm{R}(\bm{x};\sigma) &= \frac{1}{8\pi\eta r^3} \left(  \textrm{erf}\left(\frac{r}{\sigma\sqrt{2}}\right) - 4\pi r\sigma^2\Delta(\bm{x};\sigma) \right) \left(-\bm{x}\times\right)
, \label{eq:R}
\end{align}
and $(-\bm{x}\times)$ is the matrix
\begin{align}
    (-\bm{x}\times) = 
    \begin{pmatrix}
    0 & x_3 & -x_2 \\
    -x_3 & 0 & x_1 \\
    x_2 & -x_1 & 0
    \end{pmatrix}.
\end{align}
For a torque, the vorticity is given by
\begin{align}
    \bm{\omega}(\bm{x}) = \bm{P}(\bm{x};\sigma_D)\bm{T}.
\end{align}
with $\bm{P}(\bm{x};\sigma_D) = \bm{P}^{(1)}(\bm{x};\sigma_D) + \bm{P}^{(2)}(\bm{x};\sigma_D)$, where
\begin{align}
    \bm{P}^{(1)}(\bm{x};\sigma_D) &=  \frac{1}{8\pi\eta r^3} \left( -\bm{I} + \frac{3\bm{x}\bm{x}^T}{r^2} \right)\textrm{erf}\left(\frac{r}{\sigma_D\sqrt{2}}\right) \label{eq:P1} \\
    \bm{P}^{(2)}(\bm{x};\sigma_D) &= \frac{1}{2\eta r^2} \left( (\sigma_D^2 + r^2)\bm{I} - (3\sigma_D^2+r^2)\frac{\bm{x}\bm{x}^T}{r^2} \right) \Delta(\bm{x};\sigma_D)
    ,\label{eq:P2} 
\end{align}
Without presenting the details as the steps are similar those used for the force-velocity pairwise mobility, the additional pairwise mobilities are then 
\begin{align}
    \bm{M}^{\bm{\Omega F}}_{nm} & = \bm{R}(\bm{Y}_n - \bm{Y}_m;\sqrt{\sigma^2 + \sigma_D^2}) \\
    \bm{M}^{\bm{\Omega T}}_{nm} & = \frac{1}{2}\bm{P}(\bm{Y}_n - \bm{Y}_m;\sigma_D\sqrt{2}) ,
\end{align}
along with $\bm{M}^{\bm{V T}}_{nm} = \left(\bm{M}^{\bm{\Omega F}}_{mn}\right)^{\dagger}$.

As done when applying the force-velocity mobility, we split the mobility into two parts
\begin{align}
    \mathcal{M} = \widetilde{\mathcal{M}} + (\mathcal{M} - \widetilde{\mathcal{M}}), \label{eq:ewald2}
\end{align}
such that the application of $\widetilde{\mathcal{M}}$ can be evaluated rapidly on a grid while $\mathcal{M} - \widetilde{\mathcal{M}}$ is evaluated pairwise for a limited number of particle pairs.

For the grid-based computation, we solve
\begin{align}
    -\eta \nabla^2 \widetilde{\bm{u}} + \nabla \widetilde{p} &= \widetilde{\mathcal{J}}^{\dagger}[\mathcal{F}] + \widetilde{\mathcal{N}}^{\dagger}[\mathcal{T}],\\
    \nabla\cdot \widetilde{\bm{u}} &=0,
\end{align}
with $\widetilde{\mathcal{J}}^{\dagger}$ given by \eqref{eq:force_kernel_mod}, and 
\begin{align}
    \widetilde{\mathcal{N}}^{\dagger}[\mathcal{T}] = -\frac{1}{2} \sum^{N_p}_{n=1} \bm{T}^{(n)} \times \nabla  \Delta_n(\bm{x};\Sigma_D). \label{eq:torque_kernel_mod}
\end{align}
As before, the resulting particle velocities follow from applying $\widetilde{\mathcal{J}}$ to the flow field, while the angular velocities are given by 
\begin{align}
    \bm{\Omega}_m &= (\widetilde{\mathcal{N}}\left[ \bm{u} \right])_{m} = \frac{1}{2}\int_{-\infty}^{\infty} \nabla \times \tilde{\bm{u}}(\bm{x})\Delta_{m}(\bm{x}; \Sigma_D) d^3\bm{x}. \label{eq:avg_particle_rot_mod},
\end{align}
Note that for $\widetilde{\mathcal{N}}$, \textcolor{blue}{we simply replace} the Gaussian with $\sigma_D$ by one with $\Sigma_D > \sigma_D$.  No additional term involving the Laplacian is required to ensure exponential convergence.

The real space corrections are computed from the pairwise mobility relations.  They are
\begin{align}
    \bm{M}^{\bm{\Omega}\bm{F}}_{nm} - \widetilde{\bm{M}^{\bm{\Omega}\bm{F}}_{nm}} &= \bm{R}(\bm{Y}_n - \bm{Y}_m;\sqrt{\sigma^2 + \sigma_D^2}) - \bm{R}(\bm{Y}_n - \bm{Y}_m;\sqrt{\Sigma^2 + \Sigma_D^2}) \nonumber \\
    &- \frac{\sigma^2-\Sigma^2}{2} \bm{K}(\bm{Y}_n - \bm{Y}_m;\sqrt{\Sigma^2 + \Sigma_D^2}) \label{eq:correction_WF}
\end{align}
where $\bm{R}(\bm{x};\sigma)$ is provided in \eqref{eq:R} and
\begin{align}
    \bm{K}(\bm{x};\sigma) = -\frac{1}{2\sigma^2\eta}  \Delta(\bm{x}; \sigma) (-\bm{x}\times), \label{eq:V}
\end{align}
and 
\begin{align}
    \bm{M}^{\bm{\Omega T}}_{nm} - \widetilde{\bm{M}^{\bm{\Omega T}}_{nm}}= \bm{P}(\bm{Y}_n - \bm{Y}_m;\sigma_D\sqrt{2}) - \bm{P}(\bm{Y}_n - \bm{Y}_m;\Sigma_D\sqrt{2}), \label{eq:correction_WT}
\end{align}
where $\bm{P}(\bm{x}; \sigma)$ is given by \eqref{eq:P1}-\eqref{eq:P2}.  

As final point, we note that $\Sigma_D$, the width of the modified kernel appearing in $\widetilde{\mathcal{N}}$, can be set independent of $\Sigma$.  For convenience, however, in our computations below when we have torques, we take $\Sigma_D = \Sigma$.

\section{Fast FCM Algorithm} \label{sec:fastFCMalgo}
To evaluate the action of the mobility matrix using fast FCM, we must combine the grid-based FFT computation described in Section \ref{sec:gridFCM} for FCM with an efficient scheme for evaluating the pairwise correction for particles whose separation distance is within a tolerance-related cut-off radius, $R_c$.  To do this, we both sort the particles and create neighbour lists using the approach described below.  Our approach has the dual effect of ensuring efficient computation of the pairwise corrections and a localisation in memory of nearby particles, allowing for efficient spreading and interpolation to the grid.


\begin{enumerate}
\item \label{hash} \textbf{Spatial hashing}: The first step involves dividing the domain into cells in order to group particles based on their positions within the domain. This division needs to be performed only once during the initialisation of the simulation. Here, the periodic domain is divided into $m_{x}\times m_{y}\times m_{z}$ cells, where the number of cells in each direction is determined from $m_i = \textrm{max}(\textrm{int}(L_i/R_c), 3)$.  This number of cells guarantees that for all particles, any other particle within a distance $R_c$ is either in the same cell or in one of the adjacent cells.  The index for each cell is
\begin{align}
    \textrm{cell index} = x_c + (y_c + z_c\cdot m_y)\cdot m_x.
\end{align}
where $x_c = j, j=0,\dots, m_x-1$, and $y_c$ and $z_c$ are defined similarly.  At each timestep, each particle is assigned to a specific cell based on its location in the domain and the cell index is used as the particle's hash value. 

\item \label{sort} \textbf{Particle sorting and cell lists}: All particle data arrays (positions, forces, torques) are then sorted based on their hash values, which we perform using `sort by key.'  This sorting facilitates efficient memory retrieval during grid operations since particles positioned consecutively in memory will now have in common grid points to which they spread and from which they interpolate.  In addition, since particles in a cell are now contiguous in memory, the pairwise corrections for pairs in the same and adjacent cells can readily be evaluated by knowing the index of the first and the last particle in each cell.

\item \textbf{Applying} $\tilde{\mathcal{J}}^\dagger$: This step is the same as Step \ref{interpolate} in the FCM computation presented in Section \ref{sec:gridFCM} with $\mathcal{J}$ replaced by $\tilde{\mathcal{J}}^\dagger$.  
\item \textbf{Applying} $\mathcal{L}^{-1}$: This step is identical to Step \ref{solve} in Section \ref{sec:gridFCM}.

\item \textbf{Applying} $\tilde{\mathcal{J}}$: Again, this is the same as Step \ref{interpolate} in FCM computation, with $\mathcal{J}$ replaced by $\tilde{\mathcal{J}}$.  
\item \textbf{Pairwise correction}: For each particle $m$, we determine its cell index and identify the adjacent cells. Then, using the stop and start indices, we compute the distance between particle $m$ and the other particles in the same and adjacent cells.  If the distance between particles $m$ and $n$, denoted as $R_{mn}$, satisfies $R_{mn} < R_c$, we apply the pairwise correction \eqref{eq:correction_VF} using \eqref{eq:correction_WF} and \eqref{eq:correction_WT} to adjust the velocities of particles $m$ and $n$ accordingly. 
\end{enumerate}

The operation counts for applying $\tilde{\mathcal{J}}$ and $\tilde{\mathcal{J}}^{\dagger}$ will be identical to those for applying $\mathcal{J}$ and $\mathcal{J}^{\dagger}$ in FCM, as will the operation count for applying $\mathcal{L}^{-1}$.  It is important to remember, however, that the grid used for fast FCM will be smaller.  For a random dispersion of $N$ FCM particles, the operation count for the pairwise correction will be $\mathcal{O}(N\phi (R_c/a)^3)$, where $\phi$ is the volume fraction.  Thus, the savings in computation time provided by fast FCM will broadly depend on the reduction of computation time due to the reduced grid size and the increased cost of the pairwise correction.  
%

\subsection{Implementation}
In the remainder of the paper, we explore the performance of fast FCM using a C++/CUDA-based implementation that utilises the Graphics Processing Unit (GPU) to accelerate the computation.  Our CUDA implementation is freely available on GitHub: \url{https://github.com/racksa/cuFCM_demo}.

In our implementation, specific attention paid to many GPU-related considerations.  We 
carefully organise the data access pattern for quantities such as the flow field, that reside on the grid.  This limits waiting times when fetching data from memory.  We utilise the CUDA interface which offers flexibility in manipulating hardware memory through the 'shared memory' hierarchy in the GPU.  In doing so, we can take advantage of important features of GPU computing, including the ability to switch processing cores between tasks while waiting for memory communications, thereby reducing idle time.  For a more detailed summary of the CUDA-specific choices used in our implementation, please see \ref{appendix:cuda}.

\section{Error control and parameter optimisation}
We set the values of computational parameters appearing in fast FCM to ensure that particle velocities are returned efficiently to within a user specified tolerance.  These parameters are:
\begin{enumerate}
\item $M_G$, the size of the grid support of $\widetilde{\Delta}_{n}(\bm{x}; \Sigma)$ in one direction, 
\item $\Sigma/\Delta x$, the kernel width relative to the grid-spacing, 
\item $\Sigma/\sigma$, the kernel width relative to the FCM envelope width,
\item $R_c/\sigma$, the pairwise correction cutoff radius relative to the FCM envelope width.  
\end{enumerate}
The parameters $M_G$ and $\Sigma/\Delta x$ are selected to provide sufficient resolution of the grid-based computation associated with applying $\widetilde{\mathcal{M}^{\mathcal{VF}}}$.  The remaining parameters, $\Sigma/\sigma$ and $R_c/\sigma$, are set to ensure that the particle velocity error is below the specified tolerance while balancing the computational effort between the grid-based and pairwise computations.  One can see that for $\Sigma/\sigma = 1$, $R_c/\sigma$ can be rather small.  Thus, very few pairwise corrections will be required, but nothing has been gained with respect to the standard FCM computation.  At the other extreme where $\Sigma/\sigma \gg 1$, the grid size will be greatly reduced, but a large $R_c/\sigma$ will be required.  In this case, an exceedingly large number of pairwise corrections will be needed to deliver errors below an acceptable tolerance.  We seek, therefore, intermediate values of $\Sigma/\sigma$ and $R_c/\sigma$ so as to yield the optimal balance in computation time between the pairwise and grid-based computations. 

\subsection{Resolving the action of $\widetilde{\mathcal{M}^{\mathcal{VF}}}$}
We first determine the values of $M_G$ and $\Sigma/\Delta x$ that are needed to return the application of $\widetilde{\mathcal{M}^{\mathcal{VF}}}$ to within a desired tolerance.  The parameter $\Sigma/\Delta x$ controls the grid-resolution of the kernel, and $M_G$ sets the finite extent of the grid support of $\widetilde{\Delta}(\bm{x};\Sigma)$.  Hence, errors associated with $M_G$ are linked to the truncation of Gaussian kernel's tails.  As computational cost increases with $\Sigma/\Delta x$ and $M_{G}$, it is important to ensure that these are set to the lowest values possible for a given tolerance.  

\begin{figure}[h!]
    \centering
      \subfloat[\label{fig:alpha_vs_ngd}]
      {\includegraphics[width=0.5\textwidth]{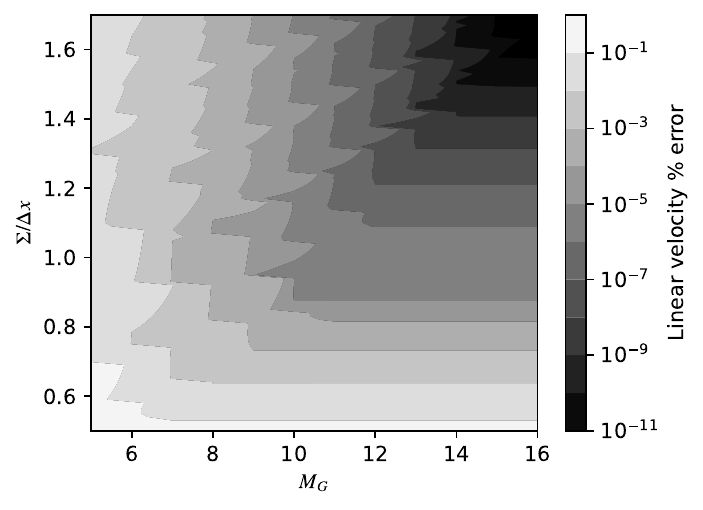}}
      \subfloat[\label{fig:rc_vs_sigmaratio}]{\includegraphics[width=0.5\textwidth]{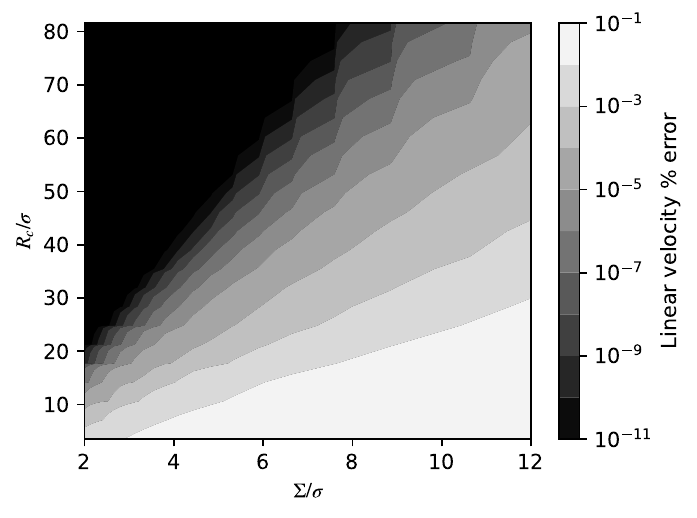}}
      \caption{Contour plots showing the fast FCM error as a function of (a) $\Sigma/\Delta x$ and $M_G$ and (b) $\Sigma/\sigma$ and $R_c/\sigma$ for a random suspension with $N=64457$, $\phi=8\%$ and $a/L=1/150$.}\label{fig:optimal_parameter}
\end{figure}

Fig. \ref{fig:optimal_parameter}\subref{fig:alpha_vs_ngd} shows the error, $\epsilon$, in applying $\widetilde{\mathcal{M}^{\mathcal{VF}}}$ for a random suspension of $N=64457$ particles with volume fraction $\phi=8\%$ in a domain of size $L\times L \times L$ for a range of $\Sigma/\Delta x$ and $M_{G}$.  To ensure any error incurred is associated with $\widetilde{\mathcal{M}^{\mathcal{VF}}}$, the correction is applied to all particle pairs by setting $R_c = L$.  Each particle $n$, is subject to a random force, $\bm{F}_n$, and the resulting velocity $\bm{V}_n$ for each $n$ is determined.  The error is given by 
\begin{align}
    \epsilon = \frac{1}{N}\sum^N_{n=1} \frac{|\bm{V}_n - \bm{U}_n|}{|\bm{U}_n|},
\end{align}
where a computation with error tolerance $10^{-15}$ was used to generate the exact velocities, $\bm{U}_n$.  We observe that $\epsilon$ decreases exponentially as $\Sigma/\Delta x$ and $M_{G}$ increase simultaneously.  We notice also that the error contours are approximately rectangular.  This indicates that minimum values of $\Sigma/\Delta x$ and $M_{G}$ are required to achieve a desired tolerance.  These values are provided in Table \ref{tab:tol_res}, where we see both $\Sigma/\Delta x$ and $M_G$ approximately double in value as $\epsilon$ decreases from $10^{-2}$ to $10^{-8}$.   

\begin{table}[h!]
    \begin{minipage}{.3\linewidth}
      \centering
        \begin{tabular}{ |c||c|c|  }
             \hline
             $\epsilon$ & $\Sigma/\Delta x$ & $M_G$\\
             \hline
             $10^{-2}$ & 0.71 & 8 \\
             $10^{-3}$ & 0.87 & 9 \\
             $10^{-4}$ & 0.99 & 10 \\
             $10^{-6}$ & 1.20 & 12 \\
             $10^{-8}$ & 1.39 & 14 \\
             \hline
        \end{tabular}
        \caption{The minimum values of $\Sigma/\Delta x$ and $M_G$ needed to achieve the error tolerance, $\epsilon$, for a random suspension with $N=64457$, $\phi=8\%$ and $a/L=1/150$.}
        \label{tab:tol_res}
    \end{minipage}%
    \hspace{.05\linewidth}
    \begin{minipage}{.65\linewidth}
      \centering
            \begin{tabular}{ |c||c|c|c|c|  }
             \hline
             $\epsilon$ & \multicolumn{4}{c|}{$\lambda_\epsilon$} \\
             \cline{2-5}
             & $\phi=0.04$ & $\phi=0.08$ & $\phi=0.16$ & $\phi=0.32$ \\
             \hline
             $10^{-2}$ & 2.31 & 2.47 & 2.63 & 2.79 \\
             $10^{-3}$ & 3.55 & 3.66 & 3.76 & 3.86 \\
             $10^{-4}$ & 4.87 & 5.08 & 5.24 & 5.43 \\
             $10^{-6}$ & 7.10 & 7.18 & 7.26 & 7.35 \\
             $10^{-8}$ & 8.53 & 8.57 & 8.63 & 8.72 \\
             \hline
        \end{tabular}
    \caption{The slope of the line $R_c/\sigma = \lambda_{\epsilon} \Sigma/\sigma$ that provides the relationship between $R_c$ and $\Sigma$ needed to achieve tolerance, $\epsilon$, for random suspensions with different $\phi$ and $a/L=1/150$.}
    \label{tab:tol_rc}
    \end{minipage} 
\end{table}

\subsection{Determining the optimal $R_c/\sigma$ and $\Sigma/\sigma$}
With values of $M_G$ and $\Sigma/\Delta x$ established for the grid-based application of $\widetilde{\mathcal{M}^{\mathcal{VF}}}$, we move to determining values for the remaining two parameters $R_c/\sigma$ and $\Sigma/\sigma$.  To do so, we must now consider the complete computation and assess how the inclusion of pairwise corrections affects the overall error.  This will provide a relationship between the values of $R_c/\sigma$ and $\Sigma/\sigma$ needed to achieve a given $\epsilon$.  Then, using parameter values lying on these curves, we can find the combination of $R_c/\sigma$ and $\Sigma/\sigma$ that minimises computational time for a specified $\epsilon$.  

Fig. \ref{fig:optimal_parameter}\subref{fig:rc_vs_sigmaratio} shows the error for the random suspension with $N=64457$ particles and volume fraction $\phi=8\%$ over a range of $R_c/\sigma$ and $\Sigma/\sigma$.  \textcolor{blue}{For all simulations performed in this paper, there is no overlapping between particles.}  The computations were performed with $\Sigma/\Delta x=2.0$ and $M_G=16$ to ensure sufficient resolution of the grid-based computation for all values of $\epsilon$.  Based on these parameter values, the number of grid points ranges from $M_x =30$ to $M_x=260$, with $M_x = M_y = M_z$ in all cases. We see that error contours can be approximated by lines that pass through the origin with slopes that increase as the error decreases.  Thus, for a given $\epsilon$, we may write $R_c/\sigma = \lambda_{\epsilon} \Sigma/\sigma$, where $\lambda_{\epsilon}$ is the slope of the line.  Table \ref{tab:tol_rc} shows the values of $\lambda_{\epsilon}$ for different values of $\epsilon$ over a range of $\phi$.  We see that for a given $\epsilon$, changing $\phi$ results in only a modest change in $\lambda_{\epsilon}$. 

While all $R_c/\sigma$ and $\Sigma/\sigma$ along these lines return values within the specified error tolerance, the computational time for different parameter values along the lines can vary widely.  At low $R_c/\sigma$ and $\Sigma/\sigma$, there will be a high computational cost incurred due to a very fine grid, while at high $R_c/\sigma$ and $\Sigma/\sigma$, the pairwise computation will dominate the cost.  Thus, the appropriate choice for $R_c/\sigma$ and $\Sigma/\sigma$ will be one along the desired tolerance curve that balances the computational costs of the grid-based and pairwise computations. 

\begin{figure}[h!]
    \centering
    {\includegraphics[width=0.8\textwidth]{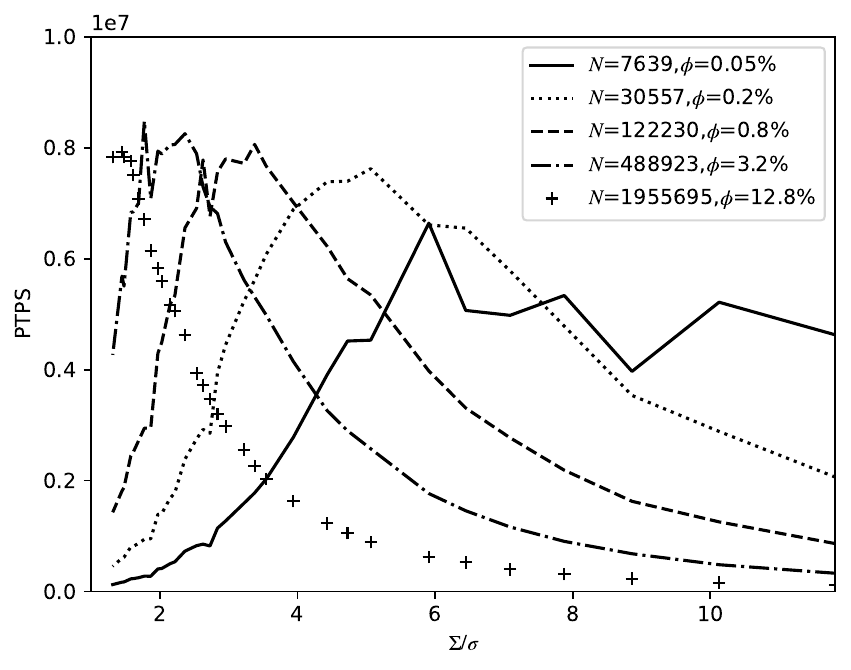}}
      \caption{The computational cost measured in particle timesteps per second (PTPS) as a function of $\Sigma/\sigma$ for random suspensions with different $\phi$.  The computations are performed with $\epsilon = 10^{-4}$ and $a/L=1/400$. Data obtained using our a single-precision CUDA implementation of fast FCM run on a single RTX 2080Ti.} \label{fig:ptps_vs_sigfac}
\end{figure}

\begin{table}[h!]
    \centering
    \begin{tabular}{ |c||c|c|  }
         \hline
         $\phi$ & optimal $\Sigma/\sigma$ & optimal $R_c/\sigma$ \\
         \hline
         0.05\% & 5.9 & 32.6  \\
         0.2\% & 5.1 & 27.9 \\
         0.8\% & 3.4 & 18.6 \\
         3.2\% & 1.8 & 9.8 \\
         12.8\% & 1.4 & 8.1 \\
         \hline
    \end{tabular}
    \caption{The optimal values of $\Sigma/\sigma$ and $R_c/\sigma$ for different volume fractions with $\epsilon = 10^{-4}$ and $a/L=1/400$.}
    \label{tab:tol_ptps}
\end{table}

Fig. \ref{fig:ptps_vs_sigfac} shows the computational cost measured in particle time-steps per second (PTPS) \cite{fiore2017rapid} as a function of $\Sigma/\sigma$ for random suspensions with different values of $\phi$.  The PTPS is the number of particles divided by the average time required to apply the mobility matrix.  Thus, high values of PTPS correspond to lower computation times.  We set the error tolerance to be $\epsilon = 10^{-4}$, and using the values from Tables \ref{tab:tol_res} and \ref{tab:tol_rc}, we have $\Sigma/\Delta x=1.00$, $M_G=10$ and $\lambda_{\epsilon}=5.5$.  We see for each $\phi$, PTPS attains a maximum value and the value of $\Sigma/\sigma$ where the peak is realised decreases as $\phi$ increases.  The values of $\Sigma/\sigma$ and $R_c/\sigma$ for which the peak values occur are given in Table \ref{tab:tol_ptps}. The peak PTPS value is approximately $7\times 10^6$ for all $\phi$ and exhibits a slight reduction at the lowest values of $\phi$.  For the most dilute case, $\phi=0.0005$, the peak occurs at $\Sigma/\sigma\approx 6$.  Thus, for this case, using fast FCM reduces the grid size by a factor of approximately $6^3$.  As the suspension becomes denser, the average particle separation decreases resulting in the peak occurring at lower values of $\Sigma/\sigma$.  Eventually, when the volume fraction becomes significantly high, the optimal computation is realised for $\Sigma/\sigma \approx 1$, and the standard FCM computation (no pairwise corrections) becomes the most efficient choice. 

\subsection{Cost comparison with standard FCM}
\begin{figure}[h!]
    \centering
      \subfloat[\label{fig:FCMperformance}]
      {\includegraphics[width=0.5\textwidth]{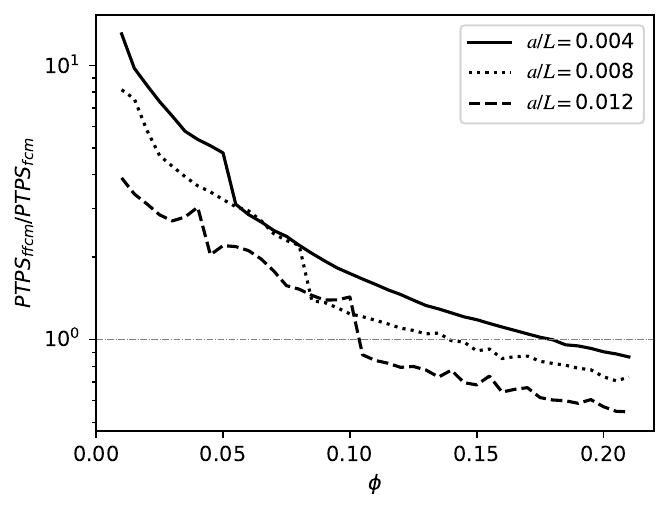}}
      \subfloat[\label{fig:FCMcrossover}]{\includegraphics[width=0.5\textwidth]{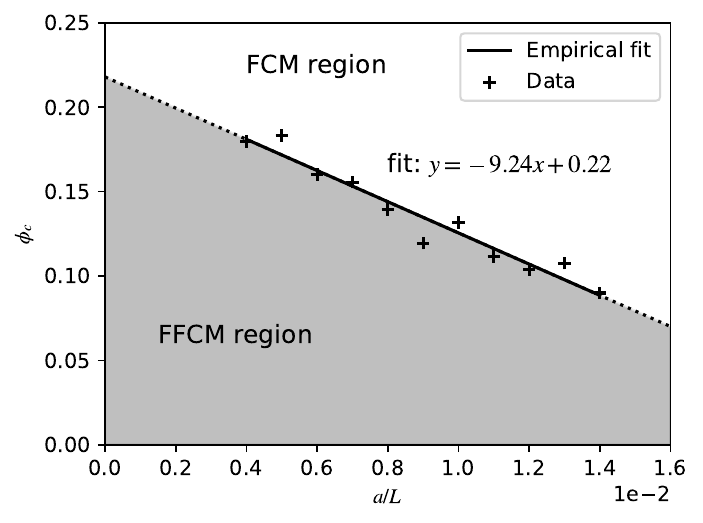}} \\
      \subfloat[\label{fig:FCMperformance2}]
      {\includegraphics[width=0.5\textwidth]{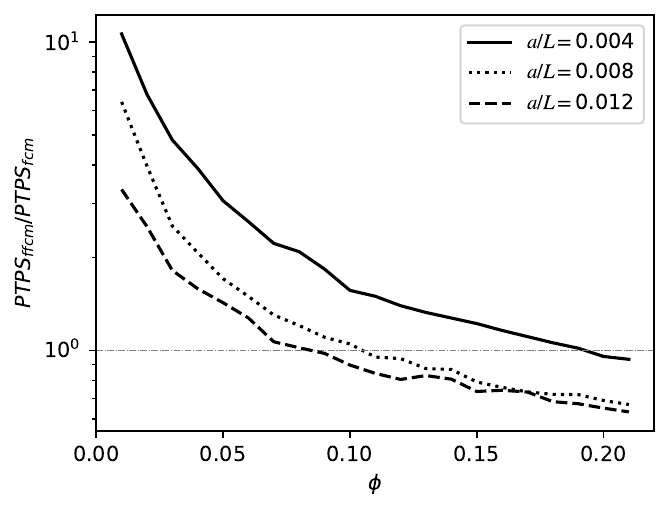}}
      \subfloat[\label{fig:FCMcrossover2}]{\includegraphics[width=0.5\textwidth]{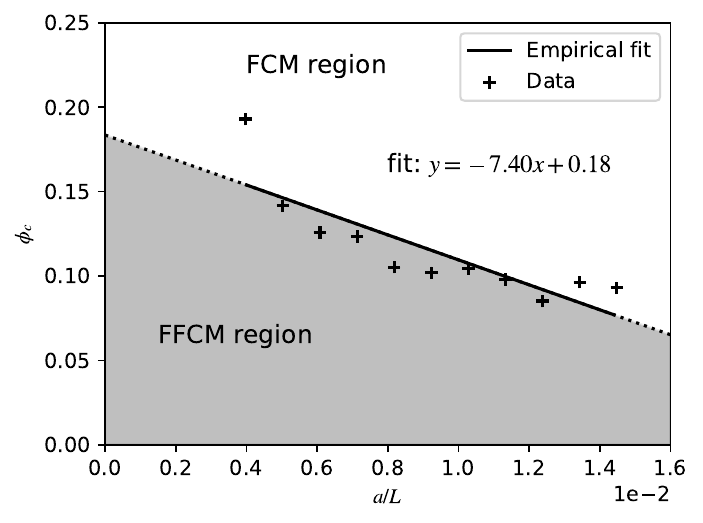}}
      \caption{(a) The ratio of PTPS for fast FCM to that for standard FCM as a function of $\phi$ for different $a/L$ and $\epsilon = 10^{-4}$.  Timings are obtained by averaging 50 computations after an initial GPU warm-up period. (b) $\phi_c$ as a function of $a/L$.  The solid line shows $\phi_c = -9.24(a/L) + 0.22$ which is determined from a linear fit to the data. 
      \textcolor{blue}{(c) The same as (a), but with $\epsilon = 10^{-6}$. (d) The same as (b), but with $\epsilon = 10^{-6}$}.}\label{fig:crossover}
\end{figure}

With the optimal parameters established, we can assess the performance of fast FCM relative to the standard FCM implementation.  To do so, we compute the PTPS for fast and standard FCM for random suspensions with different $\phi$, and determine $\phi_c$, the value of $\phi$ where the computational cost for fast FCM exceeds that of the standard implementation, i.e.  $\textrm{PTPS}_{FFCM} < \textrm{PTPS}_{FCM}$. To ensure a proper comparison, the two methods utilise the same code base and code optimisation where possible.  They both use the same FFT package and gridding subroutines, and both are run on the same device. We perform the comparison for $a/L = 0.004, 0.008$ and $0.012$, and set the tolerance to $\epsilon = 10^{-4}$.  Fig. \ref{fig:crossover}\subref{fig:FCMperformance} shows the PTPS ratio, $\textrm{PTPS}_{FFCM}/\textrm{PTPS}_{FCM}$, as a function of $\phi$ for three difference values of $a/L$.  For all cases, we have $\textrm{PTPS}_{FFCM}/\textrm{PTPS}_{FCM} > 1$ at low $\phi$ followed by an overall decrease to values $\textrm{PTPS}_{FFCM}/\textrm{PTPS}_{FCM} < 1$ as $\phi$ increases.  As discussed above, at low $\phi$ when the particles are on average further apart, the number of grid points for fast FCM is lower than that needed in standard FCM, and hence the $\textrm{PTPS}$ for fast FCM is higher.  As $\phi$ increases, the particles become more closely spaced and the number grid points for both algorithms become more comparable until reaching near equality at $\phi_c$.  The value of $\phi_c$ increases from approximately $10\%$ for $a/L = 0.012$ to approximately $18\%$ for $a/L = 0.004$.  We see, therefore, that for a fixed particle size, the computational cost of fast FCM relative to the standard computation decreases as the simulation domain increases in size.  This dependence of $\phi_c$ on $a/L$ is shown in Fig. \ref{fig:crossover}\subref{fig:FCMcrossover}.  We observe that $\phi_c$ decreases linearly, $\phi_c \approx -9.24(a/L) + 0.22$, over the range of $a/L$ that we are able to assess.  \textcolor{blue}{Similar results are obtained when the tolerance is lowered to $\epsilon = 10^{-6}$ (see Figs. \ref{fig:crossover}(c) and \ref{fig:crossover}(d))}.  For $a/L < 0.004$, the grid for standard FCM becomes too large to be accommodated in the memory of the GPU (NVidia RTX 2080 Ti), while for $a/L > 0.012$, the FCM grid is too small, resulting in idle nodes on the GPU.  For small $a/L$ but now with $\phi$ fixed, we again have that the standard FCM computation requires large $M$ to ensure $\Delta x < a$ for sufficient resolution of the particles.  This results in excessively costly grid computations, \textcolor{blue}{including very large memory overheads}, that are mitigated in fast FCM by the inclusion of the pairwise correction.  
\textcolor{blue}{Before presenting results from simulations performed using fast FCM, it is important to note that the parameter values provided here are a general guide as to how errors can be controlled and computation times minimised when using fast FCM.  Our results were compiled for simulations of random suspensions with different volume fractions performed on one type of device.  The performance of fast FCM, and hence the optimal parameter values, will depend on the particle arrangement, as well as the device on which the computation is to be performed.  Thus, users are encouraged to explore how these parameters should be tuned to optimise fast FCM for their specific computation. }



\section{Simulations using fast FCM}

In this section, we employ fast FCM as the hydrodynamic solver for simulations of rigid bodies and flexible filaments.  In these simulations, fast FCM is used in a way similar to the immersed boundary method \cite{peskin_immersed_2002} or rigid multiblob method \cite{balboa2017hydrodynamics} where the immersed body is discretised into a number of FCM particles, which then experience forces that correspond to the internal stresses experienced by the structure.  These forces can arise through constraints, or through a constitutive law that relates structure deformation to the internal stress.  As such, this allows us to examine the usefulness and performance of fast FCM as part of a larger, more complex computation.

\subsection{Sedimentation of rod suspensions}


\begin{figure}[h!]
    \centering
    \subfloat[\textrm{Illustration of the sedimentation of 2304 rods with $L=960a$.}\label{fig:single_rod}]
      {\includegraphics[width=0.4\textwidth]{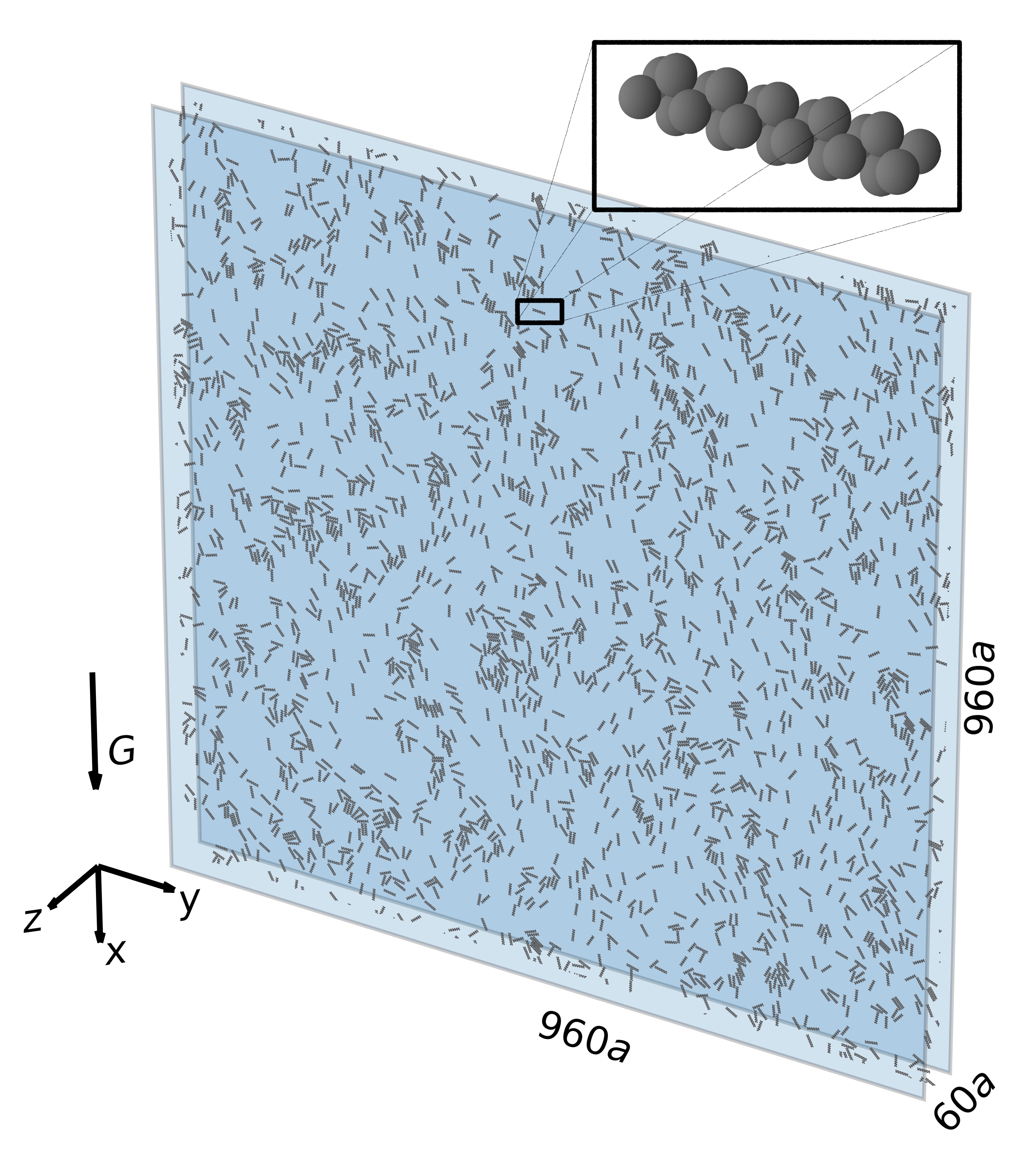}}\ \ 
      \subfloat[\label{fig:rod_vel}]
      {\includegraphics[width=0.57\textwidth] {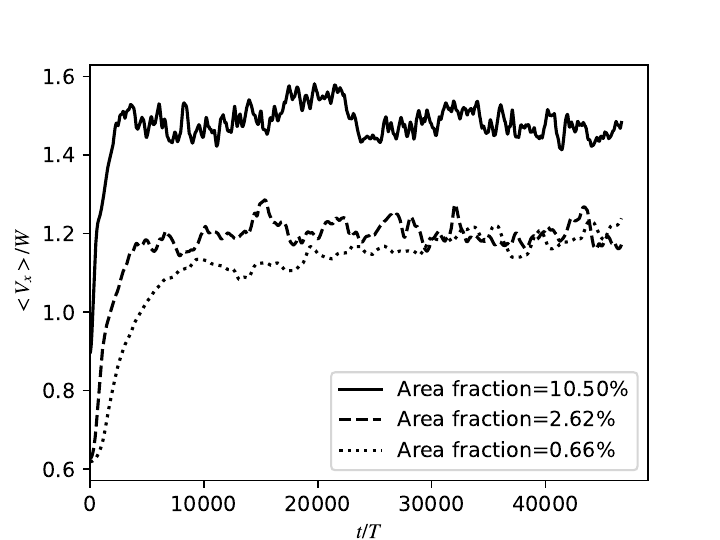}}
      \caption{(a) Image showing a snapshot from a simulation of a monolayer of $2304$ sedimenting rods in a domain of size $960a \times 960a \times 60a$.  The rods are constructed from FCM particle doublets as shown in the inset image.  (b) Average sedimentation velocity, $\langle V_x\rangle = (\sum_n \bm{V}_n\cdot\bm{\hat{x}})/N_R$, over time for simulations with $N_R = 2304$ and $L = 960a$  $(\phi=10.50\%)$ (solid line), $L = 1920a$ $(\phi=2.62\%)$ (dashed line), and $L = 3840a$ $(\phi=0.66\%)$ (dotted line).}\label{fig:rod_illustration}
\end{figure}



We first employ fast FCM to simulate the sedimentation of a planar suspension of $N_R$ rigid rods.  Following \cite{WESTWOOD2022111437}, the individual rods are constructed from 22 FCM particles stacked as alternating doublets, as shown in Figure \ref{fig:single_rod}.  As a result of this construction, the total rod length is $l=14.14a$.  The position of rod $n$ is $\bm{X}_n$, while its orientation is provided by the unit-quaternion $\bm{q}_n$.  The simulations are conducted in triply-periodic domains with dimensions $[L \times L \times 60a]$ (see again Fig. \ref{fig:rod_illustration}).

In the simulations, the rods are subject to gravity and short-ranged, inter-rod repulsive forces.  These forces are imposed on the individual FCM particles making up the rods.  For gravity, each FCM particle is subject to the constant force $\bm{F}^G = F_0\bm{\hat{x}}$, yielding the timescale $T = \eta l/F_0$.  To prevent the rods from overlapping if they collide, we apply a pairwise barrier force \cite{dance} between the FCM particles making up the rods of the form 
\begin{align} 
\bm{F}^B_{ij} = F_S\left( \frac{4a^2\chi^2 - r^2_{ij}}{4a^2(\chi^2-1)}\right)^4 \frac{\bm{r}_{ij}}{2a}
\end{align}
if $r_{ij} \leq 2 \chi a$ and $\bm{F}^B_{ij} = 0$ if $r_{nm} >2 \chi a$, where $F_S = 44F_0$, $\chi=1.1$, $\bm{r}_{ij} = \bm{Y}_i - \bm{Y}_j$, and $r_{ij} = |\bm{r}_{ij}|$.  As a result, we can write the total force and torque on rod $n$ as
\begin{align}
\bm{f}_n &= 22\bm{F}^G + \sum_{i \in \mathcal{N}_n} \bm{F}^B_i \label{eq:fn}\\
\bm{\tau}_n &= \sum_{i \in \mathcal{N}_n} (\bm{Y}_i - \bm{X}_n) \times (\bm{F}^G + \bm{F}^B_i), \label{eq:taun}
\end{align}
where $\bm{F}^B_i$ denotes the total barrier force on particle $i$ and $\mathcal{N}_n$ is the set of FCM particle indices that comprise rod $n$.  As the rods are rigid bodies, their motion is governed by, 
\begin{align}
\frac{d\bm{X}_n}{dt} &= \bm{U}_n, \label{eq:dXdt}\\
\frac{d\bm{q}_n}{dt} &= \frac{1}{2}(0, \bm{\Gamma}_n) \bullet \bm{q}_n, \label{eq:dqdt}
\end{align}
where $\bm{U}_n$ and $\bm{\Gamma}_n$ are, respectively, the translational and angular velocities of rod $n$.  The symbol $\bullet$ is the quaternion product, which for quaternions $\bm{p} = (p_0, \tilde{\bm{p}})$ and $\bm{q} = (q_0, \tilde{\bm{q}})$ is $\bm{p} \bullet \bm{q} = (p_0 q_0 - \tilde{\bm{p}}\cdot \tilde{\bm{q}}, \; p_0\tilde{\bm{q}} + q_0 \tilde{\bm{p}} + \tilde{\bm{p}} \times \tilde{\bm{q}})$.  
Computing $\bm{U}_n$ and $\bm{\Gamma}_n$ for all $n$ is done by solving 
\begin{align}
\bm{V}_i &= \bm{U}_n + \bm{\Gamma}_n \times (\bm{Y}_i - \bm{X}_n) \label{eq:ui} \\
\mathcal{V} &= \mathcal{M^{VF}}\mathcal{F} \label{eq:U} \\
\bm{f}_n &= \sum_{i \in \mathcal{N}_n} \bm{F}_i \label{eq:Fn} \\
\bm{\tau}_n &= \sum_{i \in \mathcal{N}_n} (\bm{Y}_i - \bm{X}_n) \times \bm{F}_i \label{eq:Tn}
\end{align}
where, as before $\mathcal{V} = \left[\bm{V}_1^\dagger, \bm{V}_2^\dagger, \dots, \bm{V}_N^\dagger\right]^\dagger$ and $\mathcal{F} = \left[\bm{F}_1^\dagger, \bm{F}_2^\dagger, \dots, \bm{F}_N^\dagger\right]^\dagger$ are the vectors containing the velocities and forces, respectively, on all FCM particles.  These equations enforce that conditions that the motion of individual FCM particles is consistent with the rigid body motion of the rods to which they belong.  In particular, note that $\bm{f}_n$ and $\bm{\tau}_n$ are given by \eqref{eq:fn} and \eqref{eq:taun}, respectively, and $\mathcal{F}$ must be solved for as part of the problem.

In the simulations, we discretise \eqref{eq:dXdt} and \eqref{eq:dqdt} using the second-order, geometric BDF scheme \cite{SCHOELLER2021109846} that preserves the unit length of $\bm{q}_n$ for all $n$.  The resulting algebraic equations, along with \eqref{eq:ui} - \eqref{eq:Tn}, constitute a nonlinear system, which we must solve to obtain the updated rod positions and orientations, as well as $\mathcal{F}$ that enforces the rigid body constraints.  To solve the system, we employ Broyden's method \cite{Broyden1965ACO} with an initial approximation of the Jacobian based on a diagonal mobility matrix.  At each Broyden iteration, fast FCM provides the action of $\mathcal{M^{VF}}$ on $\mathcal{F}$ to determine $\mathcal{V}$ from the current solution.  As the tolerance for Broyden's method is set to $10^{-4}$, we also require that the fast FCM return the velocities to within $\epsilon = 10^{-4}$.  At the start of the simulations, the rods are positioned on a square lattice and randomly oriented in that plane in the plane at $z = 30a$.  As $\bm{F}^G$ and $\bm{F}^B_i$ for all $i$ are in the $(x,y)$-plane and the rods are symmetric with respect to reflections about this plane, the dynamics of their positions and orientations should remain 2D as the suspension evolves.  The small out-of-plane rod motion incurred as a result of numerical error is ignored when the rod positions and orientations are advanced in time.

\begin{figure}[h!]
    \centering
      \subfloat[\label{fig:h1}]
      {\includegraphics[width=0.45\textwidth]{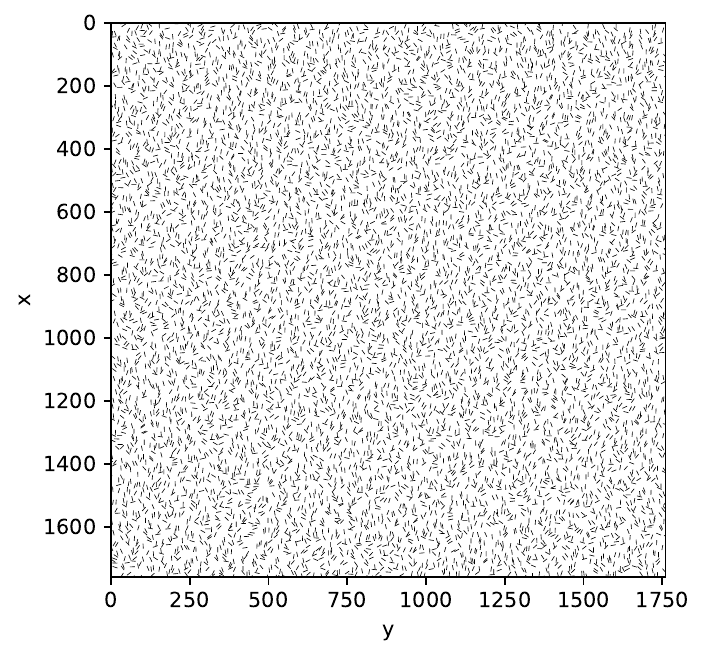}} \hfill
      \subfloat[\label{fig:h2}]
      {\includegraphics[width=0.54\textwidth] {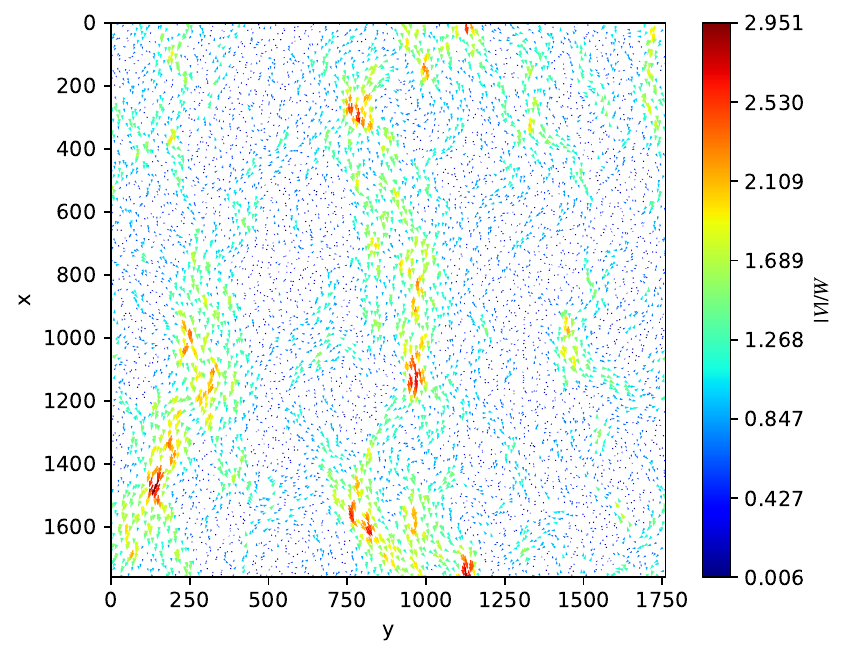}} \\
      \subfloat[\label{fig:h3}]
      {\includegraphics[width=0.45\textwidth]{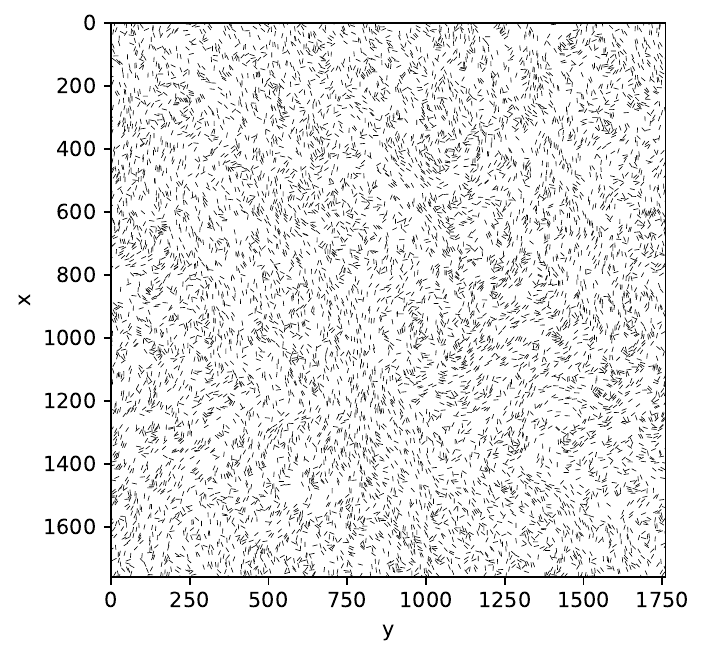}} 
      \hfill
      \subfloat[\label{fig:h4}]
      {\includegraphics[width=0.54\textwidth]{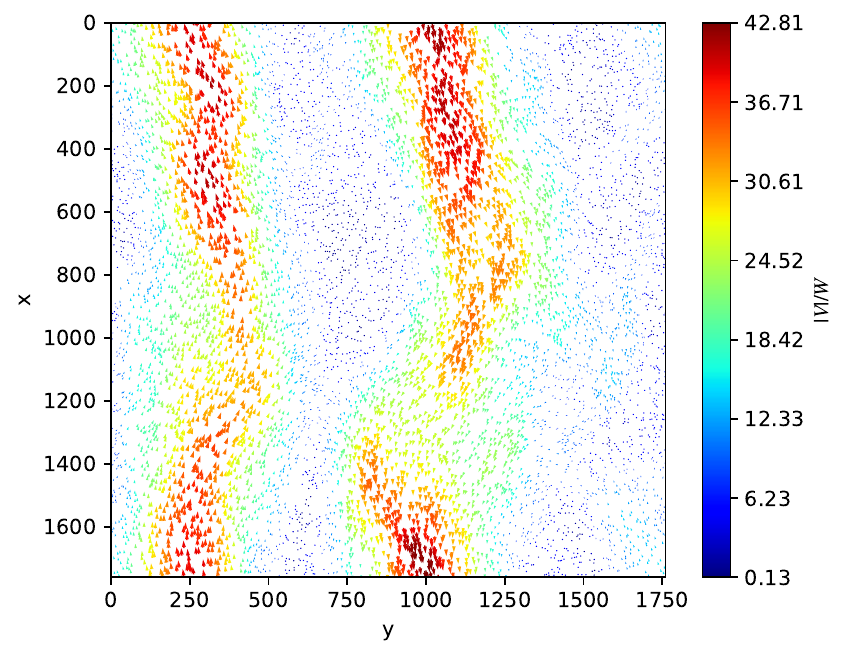}}
      \caption{(a) Snapshot of the simulation with 7744 rods at $t = 339.64T$. (b) Image showing the rod speeds at $t = 339.64T$. (c) Snapshot of the simulation with 7744 rods at $t = 4992.81T$. (b) Image showing the rod speeds at $t = 4992.81T$.} \label{fig:rod_7744}
\end{figure}

Fig. \ref{fig:rod_7744} shows the evolution of a suspension of with $N_{R} = 7744$ with an area fraction of $10.5\%$.  The simulation is run to time $t= 4992.8T$ and images of the suspension and the rod velocities at time $t = 339.6T$ and $4992.8T$ are provided.  \textcolor{blue}{Initially, when the particles have random positions and orientations, the average settling speed is below that of a single rod.}. After a short time, the rods form groups that settle rapidly, while other regions in the suspension where the rod density is lower are seen to move upwards.  \textcolor{blue}{Similar dynamics were observed in sedimentation simulations of rod suspensions in 3D periodic domains \cite{gustavsson2009gravity}}  As the rapidly falling groups encounter regions moving more slowly, they deform and eventually break up, only to reform again.  The periodic process of group formation and breaking continues over the course of the simulation. 

To understand the dynamics more quantitatively, we perform additional simulations with $N_{R} = 2304$ with a final time of $t = 46676T$.  We consider $L = 960a, 1920a,$ and $3840a$, corresponding to area fractions $10.5\%$, $2.6\%$, and $0.66\%$, respectively.  Fig. \ref{fig:rod_vel} shows the average sedimentation velocity $\langle V_x\rangle = (\sum_n \bm{V}_n\cdot\bm{\hat{x}})/N_R$ over time relative to the settling speed $W$, the sedimentation velocity of a single rod moving along its axis.  In all three cases, we see that the velocity initially increases \textcolor{blue}{from values below the settling speed of a single particle}, corresponding to the initial formation of rod clusters, as seen in Fig. \ref{fig:rod_7744}.  After this initial growth, however, the speed plateaus and fluctuates around an average value, as a result of the clusters continually breaking and reforming.  The average speed for the plateau for area fractions $0.66\%$ and $2.6\%$ are approximately equal $\langle V_x \rangle/W \approx 1.15$, while it is significantly higher for area fraction $10.5\%$, where we have $\langle V_x \rangle/W \approx 1.45$.  Additionally, the time required to reach the plateau value decreases as the area fraction increases.

\clearpage

\subsection{Arrays of filaments}
Along with simulating interacting rigid bodies, we perform simulations of arrays of interacting flexible filaments employing fast FCM as the hydrodynamic solver.  Specifically, we investigate the coordinated motion in periodic domains of an array of clamped and tethered follower-force driven filaments \cite{DeCanio2017,Ling2018Instability-drivenMicrofilaments,Westwood2021CoordinatedSurfaces,clarke2023bifurcations}.  To do so, we employ the filament model presented in \cite{SCHOELLER2021109846} and used previously in \cite{Westwood2021CoordinatedSurfaces,clarke2023bifurcations} to simulate follower-force driven filaments, which, for the sake of clarity, we describe here in the case of a single filament.

The position along the filament centerline is denoted by $\bm{Y}(s,t)$ which is a function of arc-length, $s \in [0,l)$, and time, $t$.  The filament has length $l$ and cross-sectional radius $a$.  Additionally, at each point along the filament centerline is the orthonormal basis $\{\bm{\hat{t}}(s,t),\bm{\hat{\mu}}(s,t),\bm{\hat{\nu}}(s,t)\}$, where $\bm{\hat{t}}(s,t)$ is the unit-vector tangent to the centerline,
\begin{align}
     \frac{\partial \bm{Y}}{\partial s} =\bm{\hat{t}}. \label{eq:continuous_equations3}    
\end{align}
The force and moment balances along the filament are 
\begin{align}
     \frac{\partial \bm{\Lambda}}{\partial s}+\bm{f}^H &=0, \label{eq:continuous_equations1}\\
     \frac{\partial \bm{M}}{\partial s}+\bm{\bm{\hat{t}}} \times \bm{\Lambda}+\bm{\tau}^H &=0, \label{eq:continuous_equations2}
\end{align}
where $\bm{\Lambda}(s,t)$ and $\bm{M}(s,t)$ are the internal forces and moments, respectively, on the filament cross-section, and $\bm{f}^H(s,t)$ and $\bm{\tau}^H(s,t)$ are the hydrodynamic forces and torques per unit length experienced by the filament. While the internal moments are given by a constitutive law 
\begin{equation}\label{eq:model_moment}
    \bm{M}(s,t) = K_B \left( \hat{\bm{t}} \times \frac{\partial \hat{\bm{t}}}{\partial s}\right) + K_T \left( \bm{\hat{\nu}} \cdot \frac{\partial \bm{\hat{\mu}}}{\partial s} \right) \hat{\bm{t}},
\end{equation}
where $K_B$ and $K_T$ are the bending and twist modulli, respectively, the internal forces exist to ensure \eqref{eq:continuous_equations3} is satisfied.  At $s = 0$, the filament is tethered and clamped, such that $\bm{Y}(0,t) = \bm{Y}_0$ and $\hat{\bm{t}}(0,t) = \hat{\bm{z}}$.  At $s = l$, $\bm{M}(l,t)=\bm{0}$, and 
\begin{equation}\label{eq:followerforce_BC}
    \bm{\Lambda}(l,t) = -\frac{f K_B}{l^2} \hat{\bm{t}}(l,t),
\end{equation}
where $f$ is the non-dimensional parameter controlling the magnitude of the follower force.

The filament is discretised into $N_{S}$ segments of length $\Delta l$ such that segment $i$ has position $\bm{Y}_i$ and frame $\{\bm{\hat{t}}_i,\bm{\hat{\mu}}_i,\bm{\hat{\nu}}_i\}$, which is determined from the segment quaternion, $\bm{q}_i$. Central differencing is applied to \eqref{eq:continuous_equations3}, \eqref{eq:continuous_equations1} and \eqref{eq:continuous_equations2} and after multiplying by $\Delta l$, we obtain the discrete force and moment balances,
\begin{align}
    \bm{F}_i^{C}+\bm{F}_i^H &=0,\label{eq:forcebal_discrete} \\
    \bm{T}_i^{E}+\bm{T}_i^{C}+\bm{T}_i^H &=0 \label{eq:torquebal_discrete},
\end{align}
and the discrete kinematic constraint,
\begin{equation}
    \bm{Y}_{i+1}-\bm{Y}_i-\frac{\Delta l}{2}\left(\bm{\hat{t}}_i+\bm{\hat{t}}_{i+1}\right) =0. \label{eq:inextensibility_discrete}
\end{equation}
In these expressions, $\bm{T}_i^{E}=\bm{M}_{i+1 / 2}-\bm{M}_{i-1 / 2}$ is the elastic torque, $\bm{F}_i^{C}=\bm{\Lambda}_{ i+1 / 2} - \bm{\Lambda}_{i-1 / 2}$ and $\bm{T}_i^C=(\Delta l/2) \bm{\hat{t}}_i \times (\bm{\Lambda}_{i+1 / 2}+\bm{\Lambda}_{ i-1 / 2})$ are the constraint forces and torques respectively, and $\bm{F}_i^H=\bm{f}_i^H\Delta l$ and $\bm{T}_i^H=\bm{\tau}_i^H \Delta l$ are the hydrodynamic force and torque on segment $i$. Segment velocities and angular velocities are determined by applying the FCM mobility matrix, $\mathcal{M}$, using fast FCM to the vector of hydrodynamic force and torques, 
\begin{equation}\label{eq:mobility_matrix_OG}
    \left(\begin{array}{l} \mathcal{V} \\ \mathcal{W}\end{array}\right)= \mathcal{M} \left(\begin{array}{l}\mathcal{F} \\ \mathcal{T} \end{array}\right),
\end{equation}
where $\mathcal{V}$ and $\mathcal{W}$ are, respectively, $3N_{S} \times 1$ vectors containing the components of the translational and angular velocities of all segments, and $\mathcal{F} = -\left[(\bm{F}^H_1)^\dagger, (\bm{F}_2^H)^\dagger, \dots, (\bm{F}_{N_{S}}^H)^\dagger \right]^\dagger$ and $\mathcal{T} = -\left[(\bm{T}^H_1)^\dagger, (\bm{T}_2^H)^\dagger, \dots, (\bm{T}_{N_{S}}^H)^\dagger \right]^\dagger$.  The hydrodynamic radius of each segment is $a$.

Using the translational and angular velocities, the segment positions and quaternions are obtained by integrating in time the differential-algebraic system 
\begin{align}
    \frac{d \bm{Y}_i}{d t} &=\bm{V}_i, \label{eq:continuous_equation_timeevo1} \\
    \frac{d \bm{q}_i}{dt} &= \frac{1}{2} (0,\bm{\Omega}_i) \bullet \bm{q}_i,\\ \label{eq:continuous_equation_timeevo2}
    \bm{Y}_{i+1}-\bm{Y}_i-\frac{\Delta l}{2}\left(\bm{\hat{t}}_i+\bm{\hat{t}}_{i+1}\right) &=0,
\end{align}
for each $i$.  As done in the rigid rod simulations, we discretise these equations in time using a second-order, geometric BDF scheme \cite{SCHOELLER2021109846}.  The resulting equations along with the kinematic constraints form a nonlinear system that we then solve iteratively using Broyden's method \cite{Broyden1965ACO}.  As filament interact exclusively through the surrounding fluid, the model is easily extended to many filaments by using fast FCM to couple the motion of all segments of all filaments when applying the mobilty matrix.

In our simulations, we have $N_{S} = 20$ for each filament and set the segment length $\Delta l = 2.2a$, so the total filament length is $l = 44a$.  The strength of the follower force is $f=220$.  In previous work \cite{Westwood2021CoordinatedSurfaces,clarke2023bifurcations} using the RPY tensor with a no-slip surface \cite{Swan2007SimulationBoundary} for segment mobility, this value of $f$ results in dynamics where the filament beats in a plane.  When many such filaments are in an array, their beating becomes aligned and synchronised.  Using fast FCM, we can explore how these dynamics might change under periodic boundary conditions and in the absence of a no-slip surface, where the fluid is able to flow through the array.  We examine how the filament spacing and the number of filaments affect the dynamics.

\begin{figure}[h!]
    \centering
    {\includegraphics[width=\textwidth]{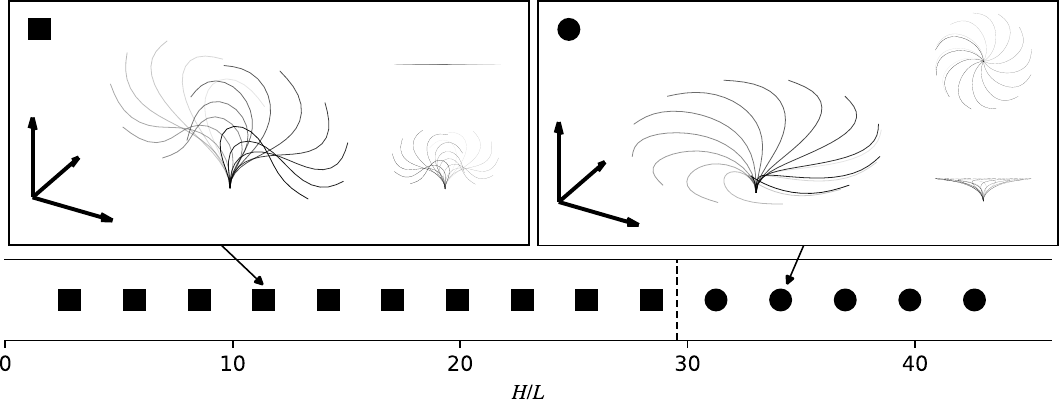}}
      \caption{Dynamics of a single filament in the periodic domain for different $H$.  Below $H = 29.54L$, the filament exhibits planar beating ($\blacksquare$), while above $H = 29.54L$ whirling (\tikzcircle[black, fill=black]{.6ex}) is observed.}\label{fig:filsingle}
\end{figure}

To begin, we first consider a single filament in the periodic domain, corresponding to an infinite array of filaments with synchronised motion.  The base of the filament is tethered at the origin and the orientation of the corresponding segment is $\bm{\hat{t}}_1 = \bm{\hat{z}}$.  The dimensions of the periodic domain are $L\times L \times H$ with $L = 2.27l$ fixed, while $H$ varies between $2.27l$ to $45.45l$.  Initially, the filament is straight and oriented vertically, i.e. $\bm{\hat{t}}_i = \bm{\hat{z}}$ for all $i$, and a random perturbation is introduced to initiate filament motion.  To ensure that we observe the asymptotically stable state, we simulate each system for at least $10000T_0$, where $T_0$ is the period of filament motion.  The resulting filament behaviour and its variation with $H$ are shown in Fig. \ref{fig:filsingle}.  We find for low values of $H < 29.54l$, we recover the planar beating observed in \cite{Westwood2021CoordinatedSurfaces,clarke2023bifurcations} with beating along one of the lattice directions. For $H > 29.54l$, however, instead of planar beating, we observe filament whirling, which is typically associated with lower values of $f$. 

\begin{figure}[h!]
    \centering
    {\includegraphics[width=\textwidth]{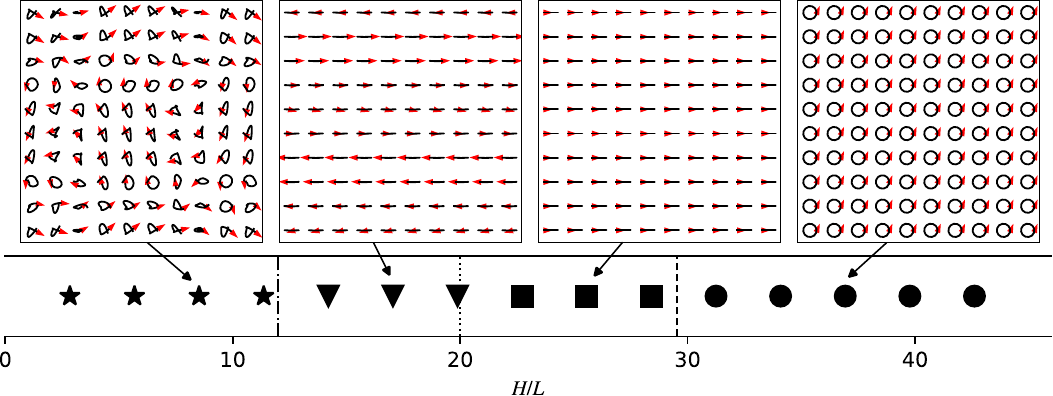}}
      \caption{Coordinated motion of 100 filaments. The black curves indicate the projection in the $xy$-plane of the filament tip trajectories over one period.  The red arrows depict the tip velocities in the final snapshot. Along with the synchronised beating ($\blacksquare$) and whirling (\tikzcircle[black, fill=black]{.6ex}) exhibited in the single filament case we also observe phase-shifted beating in rows ($\blacktriangledown$), and patches of coordinated, non-planar motion ($\bigstar$).}\label{fig:fil100}
\end{figure}

We probe the coordination further by now having $N_{F} = 100$ filaments within the periodic domain.  As these filaments can now move independently, the final state that emerges can be different from when $N_F = 1$ and synchronisation of the lattice is imposed.  To keep the filament spacing the same as the single filament case, we have $L=22.7l$ and space the filaments equally in the $x-$ and $y-$directions on a square lattice.  We vary $H$ between $2.84L$ and $45.45L$ and allow the simulations to run for more than $10000T_0$ to be confident that the final asymptotic state is reached.  The different states that emerge and the range of $H$ for which they are realised are shown in Fig. \ref{fig:fil100}.  We also provide a visual representation of the filament tip trajectories over one period as viewed from above of each state.  

We find the synchronised whirling state for $H > 29.54L$ and the synchronised beating state for $20.0L < H < 29.54L$.  These are identical to the solutions that we obtained when there was only one filament in the domain.  At low values of $H$, however, we find that two different states emerge. For $12.00L < H < 20.00L$, planar beating persists at the individual level and beating occurs in the same plane for all filaments, however, their motion is no longer synchronised across the array.  Instead, synchronised motion is observed for filaments along rows in the beat direction and phase differences develop between the different rows.  Finally, at the lowest values where $ 2.84L < H < 12.00L $, planar beating is lost and the individual filaments are found to move more erratically, often changing the direction of tip motion.  At the collective level, however, we do find synchronised movement to occur in subsections, or patches, of the array.

\subsection{Performance}

\begin{figure}[h!]
    \centering
    \subfloat[\label{fig:rod_scaling}]
      {\includegraphics[width=0.49\textwidth]{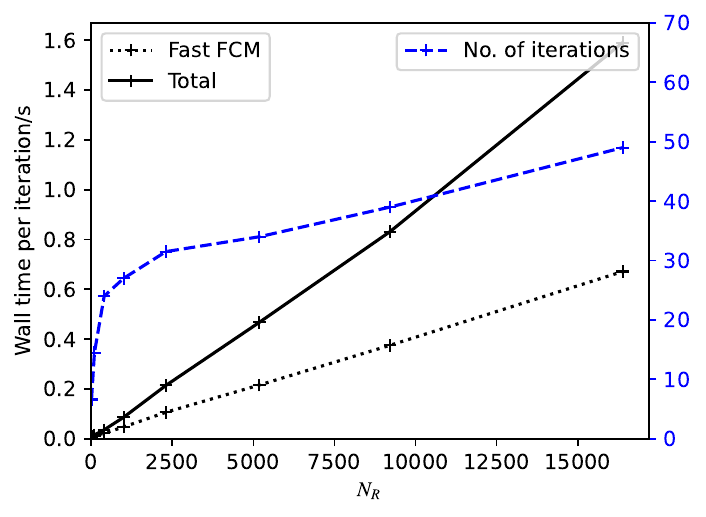}} \ \ 
      \subfloat[\label{fig:fil_scaling}]{\includegraphics[width=0.49\textwidth] {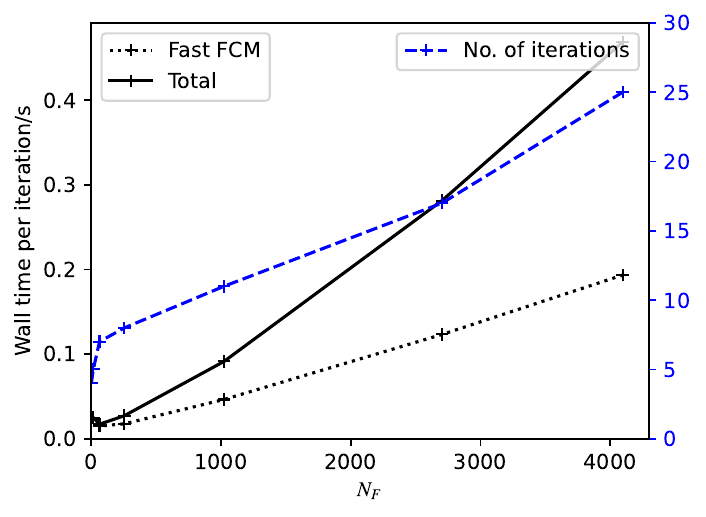}}
      \caption{Wall time per Broyden iteration and iterations per time step for (a) rod simulations with $L = 3840a$ using a grid of size $512\times512\times8$ and $R_c/\Sigma=6.0$, and (b) filament simulations in a domain of dimension $5760a \times 5760a \times 360a$ using a grid of size $256\times256\times16$ and constant $R_c/\Sigma=6.0$.  In both plots, the solid black lines show the wall time per iteration, the dotted black lines show the wall time for fast FCM per iteration, and the dashed blue lines show the average number of Broyden iterations required at each timestep.}\label{fig:scalings}
\end{figure}

Before concluding, we examine the computational performance of fast FCM for the rod and filament simulations presented above.  Fig. \ref{fig:scalings} show the average wall time required per Broyden iteration as the number of rods or filaments increases while keeping the domain size fixed.  In both cases, as expected, the time required for fast FCM to apply the mobility matrix scales linearly with the number of rods or filaments.  What we do see, again for both cases, is that with fast FCM, the computational cost of the hydrodynamics solver is reduced to approximately one third of the total cost per iteration.  This is a marked change with respect to simulations using standard FCM, where the hydrodynamic solver dominates the computational cost.  In our rod simulations, for example, portions of the computation involving rank-one updates to the Jacobian and the arithmetic operations of updating the rod positions and orientations are comparable in cost to applying the mobility matrix using fast FCM.  It is important to note that in our implemenation, updating the Jacobian and positions and orientations are both performed on the CPU rather than the GPU, and memory transfer between the CPU and the device can also affect overall wall times.  We see, therefore, that the speed and efficiency of fast FCM forces us to reevaluate the algorithms and their implementations for the nonhydrodynamic aspects of the computation.  In particular, questions as to whether more aspects of the computation can be performed using the GPU need to be considered.

Along with computational time, we also note the reduction in memory cost associated with fast FCM.  For instance, for the $N_F = 100$ filament simulations with $H=22.72L$ presented in the previous section would require a grid of approximately $1800\times1800 \times 1800$ points if standard FCM were used.  Storing the flow field and its Fourier transform on such as grid would require on the order of 280GB of RAM.  Even if such a memory requirement could be met, fetching data from memory for gridding and interpolation would compromise computational efficiency.  For fast FCM, however, the typical grid used in the $H=22.72L$ simulation contains $128\times 128\times 128$ points, requiring less than 1GB of RAM which can be easily accommodated on a single modern GPU.

\section{Conclusions and outlook}

In this paper, we have introduced fast FCM to accelerate the application of the FCM mobility matrix in periodic domains.  The method relies on decoupling the grid spacing from the FCM particle hydrodynamic radius by replacing the FCM Gaussian kernel by one with a larger width.  The resulting particle velocities from the grid-based computation are then corrected pairwise using analytical expressions to obtain values accurate to within a user specified tolerance.  We show that our specific choice of kernel ensures positive splitting, an important feature to use fast FCM for Brownian simulations, \textcolor{blue}{where we expect to compute random particle displacements using a combination of fluctuating FCM \cite{keaveny2014fluctuating,delmotte2015simulating} with the grid-based computation and the Lanczos algorithm for the pairwise corrections, as in PSE \cite{fiore2018rapid}.} We also show that fast FCM can be extended to include torques on the FCM particles.  We perform a numerical calibration of the method to obtain the optimal value of the modified envelope width and correction cut-off radius for a given tolerance.  Using our GPU based implementation on a Nvidia RTX 2080Ti, we can achieve approximately $7\times 10^6$ PTPS.  By comparing with the standard FCM computation, we find that fast FCM can, in many cases, accelerate by the application of the FCM hydrodynamic mobility matrix by an order of magnitude.  The largest gains are achieved for large-scale simulations at low volume fractions, where there are many particles that are on average far apart.  In these situations, fast FCM also avoids incurring large memory costs associated with the need to perform a large grid-based computation.  Finally, we showed that fast FCM can readily be applied as part of a larger, more complex computation for many interesting problems including rigid body suspension dynamics and the dynamics of flexible filament arrays.

There are several interesting directions in which fast FCM can be further extended.  While our general approach to accelerating FCM is similar in spirit to spectral Ewald summation \cite{lindbo2010spectrally} and Positively Split Ewald \cite{fiore2017rapid}, a key difference is that for fast FCM, the splitting is performed through a modified kernel in real space, rather than through a splitting function in Fourier space.  Thus, we are in the position to move beyond periodic domains and consider other boundary conditions such as no-slip channel walls.  Indeed, this kind of approach has been pursued \cite{graham2018microhydrodynamics,hernandez2007fast} for Stokeslets (point forces) in channels using a combination of Fourier and finite-difference methods for the grid-based solver.  Enabling fast FCM for channels would require determining mobility corrections that includes the hydrodynamic effects of the boundaries.  This would likely need to be done numerically to build up a look-up table that could then be interpolated, or fit to a convenient functional form that could then be evaluated.  We have extended fast FCM to accommodate torques applied to the FCM particles, but it remains to also include the stresslets, which are important in the context of suspension rheology \cite{yeo_simulation_2010}, or active particle dynamics \cite{delmotte2015large}.  Additionally, it will be important to show that the torque extension yields positive splitting, \textcolor{blue}{though we note that the full FCM mobility including torques is itself positive definite due to the positive definiteness of the Stokes operator, and the FCM spreading and interpolation operators being adjoint.}  Finally, a nice feature of FCM and other immersed boundary methods is the inclusion of the fluid degrees of freedom, allowing for the seamless consideration of other physics such as the interaction with chemical fields \cite{rojas2021hydrochemical}, or polymer stresses \cite{guy2014computational}.  The use of the modified kernel for the grid based computation means that we no longer compute the correct flow field.  Developing correction schemes for the flow to enable simulation in viscoelastic fluids, for example, as well as the other advancements mentioned here will form the foundation of future work along side our application of these results to interesting problems involving microscale fluid-structure interactions.

\appendix
\section{FCM mobility}\label{appendix:fcm}
In this section, we derive \eqref{eq:M_VF}, the expression for the FCM pairwise mobility matrix, $\bm{M^{VF}}_{nm}$, that relates the force on particle $m$ to the velocity on particle $n$.

The flow due to particle $m$ is given by the solution to,
\begin{align}
    -\eta \nabla^2 \bm{u}_m + \bm{\nabla} p_m &= \bm{F}_{m} \Delta_{m}(\bm{x};\sigma) \nonumber \\
    \bm{\nabla} \cdot \bm{u}_m &=0. 
\end{align}
This flow in Fourier space is 
\begin{align}
\bm{\hat{u}}_m(\bm{k}) = \hat{\mathcal{L}}^{-1}(\bm{k}) \hat{\Delta}_m(\bm{k};\sigma) \bm{F}_m,
\end{align}
where $\bm{k}$ is the wavevector and $\hat{\mathcal{L}}^{-1}(\bm{k})$ is the inverse Stokes operator in Fourier space.  Taking the inverse Fourier transform, we have 
\begin{align}
\bm{u}_m(\bm{x}) = \left[\int \hat{\mathcal{L}}^{-1}(\bm{k}) \hat{\Delta}_m(\bm{k};\sigma) \exp(i\bm{k}\cdot \bm{x}) d^3\bm{k}\right]\bm{F}_m.
\end{align}
Note that the integral in the square brackets is equivalent to $\bm{S}(\bm{x} - \bm{Y}_m; \sigma)$ in \eqref{eq:FCMflow}.  Recognising that $\Delta_{m}(\bm{x};\sigma) = \Delta(\bm{x} - \bm{Y}_m;\sigma)$, we have $\hat{\Delta}_{m}(\bm{k};\sigma) = \exp(-i\bm{k} \cdot \bm{Y}_m)\hat{\Delta}(\bm{k};\sigma)$.  Thus,
\begin{align}
\bm{u}_m(\bm{x}) = \int \hat{\mathcal{L}}^{-1}(\bm{k}) \hat{\Delta}(\bm{k};\sigma) \exp(i\bm{k}\cdot \bm{x} - \bm{Y}_m) d^3\bm{k}\bm{F}_m. \label{eq:Flow}
\end{align}
The contribution of particle $m$ to the velocity of particle $n$ is,
\begin{align}
\bm{V}_{nm} = \int \bm{u}_m(\bm{x}) \Delta_{n}(\bm{x};\sigma) d^3\bm{x},
\end{align}
which using \eqref{eq:Flow} becomes,
\begin{align}
\bm{V}_{nm} = \left[\int \left(\int \hat{\mathcal{L}}^{-1}(\bm{k}) \hat{\Delta}(\bm{k};\sigma) \exp(i\bm{k}\cdot \bm{x} - \bm{Y}_m) d^3\bm{k}\right) \Delta_{n}(\bm{x};\sigma) d^3\bm{x}\right] \bm{F}_m.
\end{align}
Performing the integration first over $\bm{x}$, and recognising that 
\begin{align}
\int \exp(i \bm{k}\cdot \bm{x})\Delta_{n}(\bm{x};\sigma) d^3\bm{x} = (2\pi)^3 \exp(i\bm{k}\cdot\bm{Y}_n) \hat{\Delta}(\bm{k};\sigma), 
\end{align}
we have
\begin{align}
\bm{V}_{nm} = \left[(2\pi)^3\int \exp(i\bm{k}\cdot(\bm{Y}_n - \bm{Y}_m)) \hat{\mathcal{L}}^{-1}(\bm{k}) \left[\hat{\Delta}(\bm{k};\sigma)\right]^2 d^3\bm{k} \right]\bm{F}_m.
\end{align}
Finally, since $(2\pi)^3 \left[\hat{\Delta}(\bm{k};\sigma)\right]^2 = (2\pi)^{-3} \exp(-k^2\sigma^2) = \hat{\Delta}(\bm{k};\sigma\sqrt{2})$,
\begin{align}
\bm{V}_{nm} = \left[\int \exp(i\bm{k}\cdot(\bm{Y}_n - \bm{Y}_m)) \hat{\mathcal{L}}^{-1}(\bm{k}) \hat{\Delta}(\bm{k};\sigma\sqrt{2}) d^3\bm{k}\right] \bm{F}_m.
\end{align}
The integral is the FCM pairwise mobility matrix, and further, we recognise that it is simply the inverse Fourier transform of the Stokes flow due to the force distribution $\Delta(\bm{x}; \sigma\sqrt{2})$.  Thus, 
\begin{align}
  \bm{M^{VF}}_{nm} &= \int \exp(i\bm{k}\cdot(\bm{Y}_n - \bm{Y}_m)) \hat{\mathcal{L}}^{-1}(\bm{k}) \hat{\Delta}(\bm{k};\sigma\sqrt{2}) d^3\bm{k} \\
  &= \bm{S}(\bm{Y}_n - \bm{Y}_m; \sigma\sqrt{2}).
\end{align}

\textcolor{blue}{
\section{Self corrections}\label{appendix:selfcorrect}
We provide below the self correction matrices for when $r = 0$:
\begin{align}
    \lim_{r\to0} \left(\bm{M}^{\bm{VF}}_{nm} - \widetilde{\bm{M}^{\bm{VF}}_{nm}}\right) = \left( \frac{1}{6\pi\eta a} - \frac{1}{6\pi\eta(\Sigma \sqrt{\pi})} + \frac{\sigma^2 - \Sigma^2}{12\eta(\Sigma\sqrt{\pi})^3} - \frac{(\sigma^2-\Sigma^2)^2}{32\eta\Sigma^{5}\pi^{3/2}}\right) \bm{I} \label{eq:correction_VF_limit}
\end{align}
\begin{align}
    \lim_{r\to0} \left(\bm{M}^{\bm{\Omega T}}_{nm} - \widetilde{\bm{M}^{\bm{\Omega T}}_{nm}}\right) = \left(\frac{1}{8\pi\eta a^3} - \frac{1}{48\eta(\Sigma\sqrt{\pi})^3}\right) \bm{I}
    \label{eq:correction_WT_limit}
\end{align}
}

\section{CUDA parallisation}\label{appendix:cuda}
We provide here the CUDA-specific parallisation strategy for the fast FCM algorithm described in Section \ref{sec:fastFCMalgo}.

\begin{enumerate}
\item For Steps \ref{hash} and \ref{sort}, we assign one thread per particle to compute the hash value and to update the cell list. To sort the particles, we employ the parallelised radix sort function in the CUDA $cub$ library.

\item For applying $\tilde{\mathcal{J}}^\dagger$ and $\tilde{\mathcal{J}}^\dagger$, we adopt a block per particle (BPP) approach.  This is one of two main approaches to particle-based gridding with CUDA, with the other being thread per particle (TPP).  For TPP, each thread is assigned one particle and is responsible for spreading the force and interpolating the velocity for that particle to the $M_G^3$ associated grid points. For BPP, each block is assigned one particle and each thread within that block is responsible for the particles spreading and interpolation for a single grid point. The advantage of BPP is that there sufficient shared memory to store the $3M_G$ values needed to reconstruct the separable kernel function at the grid points.  These values can be accessed by all threads. We use atomic operation when performing spreading to avoid race conditions (different particles attempting to write to the same grid point). For gathering, we first store the kernel weighted grid velocities in the register memory of the individual threads and, once all velocities are accounted for, a reduction operation is performed to sum the values from all threads to obtain the particle velocity. This eliminates the need for expensive atomic operations and as a result, the time for gathering is one third of that for spreading. A more detailed discussion of BPP and TPP gridding algorithms can be found in \cite{RAUL}. In our implementation of fast FCM, BPP always outperforms TPP.

\item To apply $\mathcal{L}^{-1}$, we utilise the parallelised CUDA cuFFT toolkit to perform the FFT and IFFT, assigning one thread per grid point for the computation.

\item Finally, for the pairwise correction, each thread is assigned one particle and performs the calculations to check the distance with other particles in the relevant cells, and applies the correction when required.
\end{enumerate}

\bibliography{outline}

\begin{thebibliography}{10}
\expandafter\ifx\csname url\endcsname\relax
  \def\url#1{\texttt{#1}}\fi
\expandafter\ifx\csname urlprefix\endcsname\relax\def\urlprefix{URL }\fi
\expandafter\ifx\csname href\endcsname\relax
  \def\href#1#2{#2} \def\path#1{#1}\fi

\bibitem{lauga_hydrodynamics_2009}
E.~Lauga, T.~R. Powers, The hydrodynamics of swimming microorganisms, Reports on Progress in Physics 72~(9) (2009) 096601.
\newblock \href {http://dx.doi.org/10.1088/0034-4885/72/9/096601} {\path{doi:10.1088/0034-4885/72/9/096601}}.

\bibitem{elgeti_physics_2015}
J.~Elgeti, R.~G. Winkler, G.~Gompper, Physics of microswimmers -- single particle motion and collective behavior: A review, Reports on Progress in Physics 78~(5) (2015) 056601.
\newblock \href {http://dx.doi.org/10.1088/0034-4885/78/5/056601} {\path{doi:10.1088/0034-4885/78/5/056601}}.

\bibitem{brennen_fluid_1977}
C.~E. Brennen, H.~Winet, Fluid mechanics of propulsion by cilia and flagella, Annual Review of Fluid Mechanics 9 (1977) 339--398.
\newblock \href {http://dx.doi.org/10.1146/annurev.fl.09.010177.002011} {\path{doi:10.1146/annurev.fl.09.010177.002011}}.

\bibitem{shelley2016dynamics}
M.~J. Shelley, The dynamics of microtubule/motor-protein assemblies in biology and physics, Annual review of fluid mechanics 48 (2016) 487--506.

\bibitem{mueller2010rheology}
S.~Mueller, E.~Llewellin, H.~Mader, The rheology of suspensions of solid particles, Proceedings of the Royal Society A: Mathematical, Physical and Engineering Sciences 466~(2116) (2010) 1201--1228.

\bibitem{stickel2005fluid}
J.~J. Stickel, R.~L. Powell, Fluid mechanics and rheology of dense suspensions, Annu. Rev. Fluid Mech. 37 (2005) 129--149.

\bibitem{derakhshandeh_rheology_2011}
B.~Derakhshandeh, R.~J. Kerekes, S.~G. Hatzikiriakos, C.~P.~J. Bennington, Rheology of pulp fibre suspensions: {{A}} critical review, Chemical Engineering Science 66~(15) (2011) 3460--3470.
\newblock \href {http://dx.doi.org/10.1016/j.ces.2011.04.017} {\path{doi:10.1016/j.ces.2011.04.017}}.

\bibitem{du_roure_dynamics_2017}
O.~{du Roure}, A.~Lindner, E.~N. Nazockdast, M.~J. Shelley, Dynamics of flexible fibers in viscous flows and fluids, Annual Review of Fluid Mechanics 51~(1) (2017) 539--572.
\newblock \href {http://dx.doi.org/10.1146/annurev-fluid-122316-045153} {\path{doi:10.1146/annurev-fluid-122316-045153}}.

\bibitem{de2011magnetorheological}
J.~De~Vicente, D.~J. Klingenberg, R.~Hidalgo-Alvarez, Magnetorheological fluids: a review, Soft matter 7~(8) (2011) 3701--3710.

\bibitem{moran2017phoretic}
J.~L. Moran, J.~D. Posner, Phoretic self-propulsion, Annual Review of Fluid Mechanics 49 (2017) 511--540.

\bibitem{ermak1978brownian}
D.~L. Ermak, J.~A. McCammon, Brownian dynamics with hydrodynamic interactions, The Journal of chemical physics 69~(4) (1978) 1352--1360.

\bibitem{graham2018microhydrodynamics}
M.~D. Graham, Microhydrodynamics, Brownian motion, and complex fluids, Vol.~58, Cambridge University Press, 2018.

\bibitem{brady1988stokesian}
J.~F. Brady, G.~Bossis, Stokesian dynamics, Annual review of fluid mechanics 20~(1) (1988) 111--157.

\bibitem{pozrikidis1992boundary}
C.~Pozrikidis, Boundary integral and singularity methods for linearized viscous flow, Cambridge university press, 1992.

\bibitem{balboa2017hydrodynamics}
F.~Balboa~Usabiaga, B.~Kallemov, B.~Delmotte, A.~Bhalla, B.~Griffith, A.~Donev, Hydrodynamics of suspensions of passive and active rigid particles: a rigid multiblob approach, Communications in Applied Mathematics and Computational Science 11~(2) (2017) 217--296.

\bibitem{cortez2001method}
R.~Cortez, The method of regularized stokeslets, SIAM Journal on Scientific Computing 23~(4) (2001) 1204--1225.

\bibitem{cortez2005method}
R.~Cortez, L.~Fauci, A.~Medovikov, The method of regularized stokeslets in three dimensions: analysis, validation, and application to helical swimming, Physics of Fluids 17~(3).

\bibitem{greengard1987fast}
L.~Greengard, V.~Rokhlin, A fast algorithm for particle simulations, Journal of computational physics 73~(2) (1987) 325--348.

\bibitem{tornberg2008fast}
A.-K. Tornberg, L.~Greengard, A fast multipole method for the three-dimensional stokes equations, Journal of Computational Physics 227~(3) (2008) 1613--1619.

\bibitem{hasimoto1959periodic}
H.~Hasimoto, On the periodic fundamental solutions of the stokes equations and their application to viscous flow past a cubic array of spheres, Journal of Fluid Mechanics 5~(2) (1959) 317--328.

\bibitem{lindbo2010spectrally}
D.~Lindbo, A.-K. Tornberg, Spectrally accurate fast summation for periodic stokes potentials, Journal of Computational Physics 229~(23) (2010) 8994--9010.

\bibitem{fiore2017rapid}
A.~M. Fiore, F.~Balboa~Usabiaga, A.~Donev, J.~W. Swan, Rapid sampling of stochastic displacements in brownian dynamics simulations, The Journal of chemical physics 146~(12).

\bibitem{sierou2001accelerated}
A.~Sierou, J.~F. Brady, Accelerated stokesian dynamics simulations, Journal of fluid mechanics 448 (2001) 115--146.

\bibitem{fiore2019fast}
A.~M. Fiore, J.~W. Swan, Fast stokesian dynamics, Journal of Fluid Mechanics 878 (2019) 544--597.

\bibitem{peskin_immersed_2002}
C.~S. Peskin, The immersed boundary method, Acta Numerica 11 (2002) 479--517.
\newblock \href {http://dx.doi.org/10.1017/S0962492902000077} {\path{doi:10.1017/S0962492902000077}}.

\bibitem{bringley2008validation}
T.~T. Bringley, C.~S. Peskin, Validation of a simple method for representing spheres and slender bodies in an immersed boundary method for stokes flow on an unbounded domain, Journal of Computational Physics 227~(11) (2008) 5397--5425.

\bibitem{maxey_localized_2001}
M.~R. Maxey, B.~K. Patel, Localized force representations for particles sedimenting in {{Stokes}} fow, International Journal of Multiphase Flow 27 (2001) 1603--1626.
\newblock \href {http://dx.doi.org/10.1016/S0301-9322(01)00014-3} {\path{doi:10.1016/S0301-9322(01)00014-3}}.

\bibitem{lomholt_force-coupling_2003}
S.~Lomholt, M.~R. Maxey, Force-coupling method for particulate two-phase flow: {{Stokes}} flow, Journal of Computational Physics 184 (2003) 381--405.
\newblock \href {http://dx.doi.org/10.1016/s0021-9991(02)00021-9} {\path{doi:10.1016/s0021-9991(02)00021-9}}.

\bibitem{yeo_simulation_2010}
K.~Yeo, M.~R. Maxey, Simulation of concentrated suspensions using the force-coupling method, Journal of Computational Physics 229~(6) (2010) 2401--2421.
\newblock \href {http://dx.doi.org/10.1016/j.jcp.2009.11.041} {\path{doi:10.1016/j.jcp.2009.11.041}}.

\bibitem{hernandez2007fast}
J.~P. Hern{\'a}ndez-Ortiz, J.~J. de~Pablo, M.~D. Graham, Fast computation of many-particle hydrodynamic and electrostatic interactions in a confined geometry, Physical review letters 98~(14) (2007) 140602.

\bibitem{keaveny2014fluctuating}
E.~E. Keaveny, Fluctuating force-coupling method for simulations of colloidal suspensions, Journal of Computational Physics 269 (2014) 61--79.

\bibitem{beenakker1986ewald}
C.~Beenakker, Ewald sum of the rotne--prager tensor, The Journal of chemical physics 85~(3) (1986) 1581--1582.

\bibitem{WESTWOOD2022111437}
T.~A. Westwood, B.~Delmotte, E.~E. Keaveny, \href{https://www.sciencedirect.com/science/article/pii/S0021999122004995}{A generalised drift-correcting time integration scheme for brownian suspensions of rigid particles with arbitrary shape}, Journal of Computational Physics 467 (2022) 111437.
\newblock \href {http://dx.doi.org/https://doi.org/10.1016/j.jcp.2022.111437} {\path{doi:https://doi.org/10.1016/j.jcp.2022.111437}}.
\newline\urlprefix\url{https://www.sciencedirect.com/science/article/pii/S0021999122004995}

\bibitem{dance}
S.~L. Dance, E.~Climent, M.~R. Maxey, \href{https://doi.org/10.1063/1.1637349}{Collision barrier effects on the bulk flow in a random suspension}, Physics of Fluids 16~(3) (2004) 828--831.
\newblock \href {http://arxiv.org/abs/https://doi.org/10.1063/1.1637349} {\path{arXiv:https://doi.org/10.1063/1.1637349}}, \href {http://dx.doi.org/10.1063/1.1637349} {\path{doi:10.1063/1.1637349}}.
\newline\urlprefix\url{https://doi.org/10.1063/1.1637349}

\bibitem{SCHOELLER2021109846}
S.~F. Schoeller, A.~K. Townsend, T.~A. Westwood, E.~E. Keaveny, \href{https://www.sciencedirect.com/science/article/pii/S0021999120306203}{Methods for suspensions of passive and active filaments}, Journal of Computational Physics 424 (2021) 109846.
\newblock \href {http://dx.doi.org/https://doi.org/10.1016/j.jcp.2020.109846} {\path{doi:https://doi.org/10.1016/j.jcp.2020.109846}}.
\newline\urlprefix\url{https://www.sciencedirect.com/science/article/pii/S0021999120306203}

\bibitem{Broyden1965ACO}
C.~G. Broyden, A class of methods for solving nonlinear simultaneous equations, Mathematics of Computation 19 (1965) 577--593.

\bibitem{gustavsson2009gravity}
K.~Gustavsson, A.-K. Tornberg, Gravity induced sedimentation of slender fibers, Physics of fluids 21~(12).

\bibitem{DeCanio2017}
G.~De~Canio, E.~Lauga, R.~E. Goldstein, {Spontaneous oscillations of elastic filaments induced by molecular motors}, Journal of the Royal Society Interface 14~(136).
\newblock \href {http://dx.doi.org/10.1098/rsif.2017.0491} {\path{doi:10.1098/rsif.2017.0491}}.

\bibitem{Ling2018Instability-drivenMicrofilaments}
F.~Ling, H.~Guo, E.~Kanso, {Instability-driven oscillations of elastic microfilaments}, Journal of the Royal Society Interface 15~(149).
\newblock \href {http://dx.doi.org/10.1098/rsif.2018.0594} {\path{doi:10.1098/rsif.2018.0594}}.

\bibitem{Westwood2021CoordinatedSurfaces}
T.~A. Westwood, E.~E. Keaveny, {Coordinated motion of active filaments on spherical surfaces}, Physical Review Fluids 6~(12).
\newblock \href {http://dx.doi.org/10.1103/PhysRevFluids.6.L121101} {\path{doi:10.1103/PhysRevFluids.6.L121101}}.

\bibitem{clarke2023bifurcations}
B.~Clarke, Y.~Hwang, E.~Keaveny, Bifurcations and nonlinear dynamics of the follower force model for active filaments, arXiv preprint arXiv:2309.06294.

\bibitem{Swan2007SimulationBoundary}
J.~W. Swan, J.~F. Brady, {Simulation of hydrodynamically interacting particles near a no-slip boundary}, Physics of Fluids 19~(11).
\newblock \href {http://dx.doi.org/10.1063/1.2803837} {\path{doi:10.1063/1.2803837}}.

\bibitem{delmotte2015simulating}
B.~Delmotte, E.~E. Keaveny, Simulating brownian suspensions with fluctuating hydrodynamics, The Journal of chemical physics 143~(24).

\bibitem{fiore2018rapid}
A.~M. Fiore, J.~W. Swan, Rapid sampling of stochastic displacements in brownian dynamics simulations with stresslet constraints, The Journal of chemical physics 148~(4).

\bibitem{delmotte2015large}
B.~Delmotte, E.~E. Keaveny, F.~Plourabou{\'e}, E.~Climent, Large-scale simulation of steady and time-dependent active suspensions with the force-coupling method, Journal of Computational Physics 302 (2015) 524--547.

\bibitem{rojas2021hydrochemical}
F.~Rojas-P{\'e}rez, B.~Delmotte, S.~Michelin, Hydrochemical interactions of phoretic particles: a regularized multipole framework, Journal of Fluid Mechanics 919 (2021) A22.

\bibitem{guy2014computational}
R.~D. Guy, B.~Thomases, Computational challenges for simulating strongly elastic flows in biology, in: Complex Fluids in Biological Systems: Experiment, Theory, and Computation, Springer, 2014, pp. 359--397.

\bibitem{RAUL}
R.~P. Palaez, \href{https://github.com/RaulPPelaez/UAMMD}{Complex fluids in the Gpu era}, 2022.
\newline\urlprefix\url{https://github.com/RaulPPelaez/UAMMD}

\end{thebibliography}

\end{document}